\documentclass[twocolumn,]{aastex7}

\usepackage{multirow}
\usepackage{graphicx}	
\usepackage{amsmath}	
\defcitealias{2019NatAs.tmp..258I}{I19}
\defcitealias{2021MNRAS.507.1127K}{K21}
\defcitealias{2012ApJ...751....6D}{DC12}
\defcitealias{2025MNRAS.537.2752K}{KI25}
\usepackage{caption}
\usepackage{subcaption}

\DeclareRobustCommand{\VAN}[3]{#2}
\let\VANthebibliography\thebibliography
\def\thebibliography{\DeclareRobustCommand{\VAN}[3]{##3}\VANthebibliography}

\def\Pmem{{$\mathrm{P}_{\mathrm{mem}}$}}
\def\fehpgs{{$\mathrm{[Fe/H]_{PGS}}$}}

\def\g{{\rm\,g}}

\def \r {r$^{1/4}$ }

\def\s{\ifmmode \widetilde \else \~\fi}
\def\={\overline}

\def\spose#1{\hbox to 0pt{#1\hss}}

\def\lta{\mathrel{\spose{\lower 3pt\hbox{$\mathchar"218$}}
     \raise 2.0pt\hbox{$\mathchar"13C$}}}
\def\gta{\mathrel{\spose{\lower 3pt\hbox{$\mathchar"218$}}
     \raise 2.0pt\hbox{$\mathchar"13E$}}}
\def\Dt{\spose{\raise 1.5ex\hbox{\hskip3pt$\mathchar"201$}}}    
\def\dt{\spose{\raise 1.0ex\hbox{\hskip2pt$\mathchar"201$}}}    

\def\r{${\rm r^{1/4}}$}

\def\dotsfill{\leaders\hbox to 1em{\hss.\hss}\hfill}
\def\sun{\odot}
\def\earth{\oplus}

\def\FeH{{\rm[Fe/H]}}

\def\FeHPr{{\rm\FeH_\mathrm{PGS}}}

\begin{document}

\title[Pristine View of Extended Globular Cluster Structure]{Constructing a Pristine View of Extended Globular Cluster Structure}

\author[orcid=0000-0003-1980-8838,sname='Kuzma']{Pete B. Kuzma}
\altaffiliation{JSPS International Research Fellow}
\affiliation{National Astronomical Observatory of Japan, 2-21-1 Osawa, Mitaka, Tokyo 181-8588, Japan}
\affiliation{Institute for Astronomy, University of Edinburgh, Royal Observatory, Blackford Hill, Edinburgh, EH9 3HJ, UK}

\email[show]{pete.kuzma@nao.ac.jp}  

\author[sname='Ishigaki']{Miho N. Ishigaki} 
\affiliation{National Astronomical Observatory of Japan, 2-21-1 Osawa, Mitaka, Tokyo 181-8588, Japan}
\affiliation{Graduate Institute for Advanced Studies, SOKENDAI, 2-21-1 Osawa, Mitaka, Tokyo 181-8588, Japan}
\affiliation{Kavli Institute for the Physics and Mathematics of the Universe (WPI), The University of Tokyo Institutes for Advanced Study, The University of Tokyo, Kashiwa, Chiba 277-8583, Japan}
\email{miho.ishigaki@nao.ac.jp}

\author[sname='Kirihara']{Takanobu Kirihara}
\affiliation{Kitami Institute of Technology, 165, Koen-cho, Kitami, Hokkaido 090-8507, Japan}
\email{tkirihara@mail.kitami-it.ac.jp}

\author[sname='Ogami']{Itsuki Ogami}
\affiliation{National Astronomical Observatory of Japan, 2-21-1 Osawa, Mitaka, Tokyo 181-8588, Japan}
\affiliation{The Institute of Statistical Mathematics, 10-3 Midoricho, Tachikawa, Tokyo 190-8562, Japan}
\email{itsuki.ogami@nao.ac.jp}

\begin{abstract}

Globular Clusters (GCs) displaying extended structures are becoming increasingly ubiquitous in the Milky Way (MW). Despite their low surface brightness, which makes disentangling the true structure from the MW overwhelming difficult, the increasing availability of multi-dimensional data sets has allowed for new detections of extended GC structure. This work utilises the Pristine-Gaia-Synthetic catalogue released as part of the Pristine Surveys first data release to search for tidally stripped stars in the peripheries of MW GCs. Pristine provides photometric [Fe/H] measurements based on CaHK-band photometry. Using unsupervised machine learning techniques, we provided lists of extra-tidal stars for 30 GCs, one of the largest surveys of its kind. We find that (1) 22 GCs that passed our quality cut have extended structure within 5 deg from the cluster centers of which six are new tentative detections, (2) four of those GCs exhibit diffuse envelope-like extra-tidal features, while the remaining GCs exhibit tidal tail-like structures. We measure the position angles of the extended structures, find broad consistency between the position angles and the GC orbits, and discuss our results concerning N-body models. This work demonstrates the effectiveness of adding photometric metallicities to the multi-dimensional search of extended tidal structure and how the upcoming multi-object spectrographs will be crucial for exploring GC peripheries in the coming years.

\end{abstract}



\section{Introduction}
Globular Clusters (GCs) are more than just a simple spherical distribution of stars. These ancient stellar populations litter the Milky Way (MW) halo, with the latest estimates placing the number of GCs at least 178 \cite[][]{2019A&A...630L...4M,2025arXiv250314657M}. Furthermore, they can show complex stellar populations, with most GCs possessing unique light element abundance variations such as anti-correlations in Na-O and Mg-Al, and a small but not insignificant subset even possessing variations in [Fe/H] and heavier elements \citep[see the following reviews][and references therein]{1994PASP..106..553K,2004ARA&A..42..385G,2012A&ARv..20...50G,2018ARA&A..56...83B,2019A&ARv..27....8G,2022Univ....8..359M}. These properties suggest a complex formation history for GCs that may represent the conditions of the early universe. Beyond the chemistry of these GCs, their kinematics are also far from simple. For example, the proper motions provided by Gaia \citep[][]{2016A&A...595A...1G} have allowed the exploration of the origins of the MW GC system. Conserved energy spaces, such as orbital actions, have been used to tag GCs to different proposed accretion events \citep[e.g.,][]{malhan_global_2022,2022MNRAS.513.4107C,2023RAA....23a5013S}. While many recent studies have shown that orbital actions are not conserved during merger events, \citep[][]{2017A&A...604A.106J,2024A&A...690A.136M,2023A&A...673A..86P}, the GC chemical abundances paired with orbital properties continue to offer insights into the Galaxy's accretion history \citep[][]{2025A&A...693A.155P}. These exciting findings have all been established by exploring the inner regions of GCs, though there is more to be found in their peripheries and beyond.


Studying the extended structure of GCs can shed light on many facets of GCs that the central regions of GCs can not assist with. For example, modelling of the extended structure can be used to infer the shape of the Galactic halo out to the distance of targeted GC \citep[][]{2014ApJ...795...94B,2016ApJ...833...31B,2025arXiv250213236T}, assess the existence of GCs embedded in dark matter subhalos \citep[][]{penarrubia_stellar_2017,2021ApJ...922..104C,2022ApJ...935...14C,2025MNRAS.537.1807A} or assess the black hole populations within the GC \citep[][]{2021NatAs...5..957G,2025MNRAS.538..454R}. Furthermore, some GCs show signs of large extended tidal structure: diffuse, faint stellar structures whose origins are not completely understood. Since the seminal discovery of tidal tails around Pal 5 in the Sloan Digital Sky Survey \citep[][]{odenkirchen_detection_2001}, the number of GCs with extended tidal structure has slowly grown as our technologies and instruments provided higher quality data and new novel techniques were developed \citep[e.g.,][]{2000A&A...359..907L,2010A&A...522A..71J,2018MNRAS.473.2881K,2019MNRAS.486.1667C,2019MNRAS.485.1029P,2019ApJ...884..174G,2019MNRAS.488.1535P,2020MNRAS.495.2222S,2020MNRAS.499.2157C,2021MNRAS.507.1127K,2024arXiv240513498M,2024AJ....167..279P,2024A&A...683A.151P}. As a result, the nature of the GC system of the MW, and their relationship with not only accreted dwarf galaxies but to the numerous stellar streams and substructures newly uncovered in the stellar halo, was placed under the spotlight \citep[e.g.,][]{2018MNRAS.474.2479B,2020A&A...637L...2P,2025NewAR.10001713B}. However, studying these regions is not without its difficulties.

\begin{deluxetable*}{lrrrrrrrr}
\tablewidth{0pt}
\tablecaption{Parameters of the GCs studied in this work. Columns are as follows: (1) Name of GC, (2) Right ascension (J2000), (3) Declination (J2000), (4) Proper motion in the R. A. direction, (5) Proper motion in the Dec. direction \citep[][]{2021MNRAS.505.5978V}, (6) [Fe/H] measurements \citep[][]{2009AA...508..695C}, (7) Heliocentric distance \citep[][]{baumgardt_accurate_2021}, (8) tidal radius \citep[][]{de_boer_globular_2019} and (9) Jacobi radius \citep[][]{2018MNRAS.474.2479B}.\label{tab:GC_list}}
\tablehead{
\colhead{Cluster} & \colhead{R. A.} & \colhead{Dec.} & \colhead{$\mu_\alpha$} & \colhead{$\mu_\delta$} & \colhead{[Fe/H]} & \colhead{$R_\odot$} & \colhead{$r_t$} & \colhead{$r_J$} \\
\colhead{} & \colhead{(HH:MM:SS)} & \colhead{(DD:MM:SS)} & \colhead{(mas yr$^{-1}$)} & \colhead{(mas yr$^{-1}$)} & \colhead{(dex)} & \colhead{(kpc)} & \colhead{(pc)} & \colhead{(pc)}
}
\startdata
NGC 104 & 00:24:05.67 & -72:04:52.6 & 5.237 & -2.524 & -0.72 & 4.5 & 52.49 & 137.45\\
NGC 288 & 00:52:45.24 & -26:34:57.4 & 4.252 & -5.641 & -1.32 & 8.9 & 46.31 & 76.43\\
NGC 362 & 01:03:14.26 & -70:50:55.6 & 6.730 & -2.535 & -1.26 & 8.6 & 26.08 & 112.06\\
NGC 1261 & 03:12:16.21 & -55:12:58.4 & 1.632 & -2.038 & -1.27 & 16.3 & 37.81 & 146.38\\
NGC 1851 & 05:14:06.76 & -40:02:47.6 & 2.120 & -0.589 & -1.18 & 12.1 & 31.92 & 166.46\\
NGC 1904 & 05:24:10.60 & -24:31:28.1 & 2.467 & -1.573 & -1.60 & 12.9 & 38.32 & 153.79\\
NGC 2298 & 06:48:59.41 & -36:00:19.1 & 3.316 & -2.186 & -1.92 & 10.8 & 30.05 & 81.27\\
NGC 2808 & 09:12:03.10 & -64:51:48.6 & 1.005 & 0.274 & -1.14 & 9.6 & 28.91 & 176.87\\
NGC 3201 & 10:17:36.82 & -46:24:44.9 & 8.324 & -1.991 & -1.59 & 4.9 & 57.10 & 83.46\\
NGC 4590 & 12:39:27.98 & -26:44:38.6 & -2.752 & 1.762 & -2.23 & 10.3 & 50.77 & 88.54\\
NGC 5272 & 13:42:11.62 & 28:22:38.2 & -0.142 & -2.647 & -1.50 & 10.2 & 77.55 & 158.64\\
NGC 5466 & 14:05:27.29 & 28:32:04.0 & -5.414 & -0.805 & -1.98 & 16.1 & 103.60 & 105.03\\
NGC 5897 & 15:17:24.50 & -21:00:37.0 & -5.427 & -3.438 & -1.90 & 12.5 & 41.94 & 62.99\\
NGC 5904 & 15:18:33.22 & 02:04:51.7 & 4.078 & -9.854 & -1.29 & 7.5 & 56.56 & 100.52\\
NGC 6101 & 16:25:48.12 & -72:12:07.9 & 1.757 & -0.223 & -1.98 & 15.4 & 27.51 & 78.02\\
NGC 6205 & 16:41:41.23 & 36:27:35.5 & -3.164 & -2.588 & -1.53 & 7.1 & 35.69 & 113.26\\
NGC 6218 & 16:47:14.18 & -01:56:54.7 & -0.141 & -6.802 & -1.37 & 4.8 & 29.35 & 48.74\\
NGC 6341 & 17:17:07.38 & 43:08:09.4 & -4.925 & -0.536 & -2.31 & 8.3 & 23.29 & 103.5\\
NGC 6362 & 17:31:54.98 & -67:02:54.0 & -5.510 & -4.750 & -0.99 & 7.6 & 38.31 & 44.76\\
NGC 6397 & 17:40:42.09 & -53:40:27.6 & 3.285 & -17.621 & -2.02 & 2.3 & 32.16 & 45.96\\
NGC 6541 & 18:08:02.36 & -43:42:53.6 & 0.349 & -8.843 & -1.81 & 7.5 & 25.98 & 43.90\\
NGC 6584 & 18:18:37.60 & -52:12:56.8 & -0.053 & -7.185 & -1.50 & 13.5 & 30.82 & 69.47\\
NGC 6681 & 18:43:12.76 & -32:17:31.6 & 1.458 & -4.688 & -1.62 & 9.4 & 23.77 & 22.22\\
NGC 6752 & 19:10:52.10 & -59:59:04.4 & -3.170 & -4.043 & -1.54 & 4.1 & 32.19 & 63.14\\
NGC 6809 & 19:39:59.71 & -30:57:53.1 & -3.420 & -9.269 & -1.94 & 5.4 & 29.55 & 46.50\\
NGC 6934 & 20:34:11.37 & 07:24:16.1 & -2.636 & -4.667 & -1.47 & 15.6 & 26.12 & 103.08\\
NGC 6981 & 20:53:27.70 & -12:32:14.3 & -1.233 & -3.290 & -1.42 & 16.7 & 30.55 & 90.82\\
NGC 7078 & 21:29:58.33 & 12:10:01.2 & -0.643 & -3.763 & -2.37 & 10.4 & 39.57 & 158.56\\
NGC 7089 & 21:33:27.02 & -00:49:23.7 & 3.518 & -2.145 & -1.65 & 11.5 & 40.18 & 149.59\\
NGC 7099 & 21:40:22.11 & -23:10:47.5 & -0.694 & -7.271 & -2.27 & 8.1 & 31.35 & 67.86\\
\enddata
\end{deluxetable*}

Many reports of tidal tails and extended tidal structure around GCs have come in the past decade, indicating the troublesome nature of exploring large data sets for these features. The limitations of studying outer regions of GCs lie with the fact that typically, they are populated with lower-mass stars \citep[e.g.,][]{2004AJ....127.2753D,kupper_tidal_2010,2023ApJ...946..104W}. In turn, reaching these stars typically invites more MW field contamination, creating difficulties in identifying GC stars. Extra dimensions in the data beyond photometry are required to enable a true disentanglement of GCs stars from the overwhelming MW field populations. The astrometry of Gaia revolutionised our ability to search for tidally stripped stars surrounding GCs. On-sky proper motions provided a crucial extra dimension that can remove field contamination. However, there are dimensions where Gaia is lacking. The third data release of Gaia provides [Fe/H] and line-of-sight velocities \citep[][]{2023AA...674A...1G}, two crucial additional dimensions, but only for the survey's brighter stars. As a result, identifying extra-tidal stars in the peripheries relies primarily on proper motions. Identifying individual stars in extended structures using line-of-sight velocities is primarily done through spectroscopy, as multi-object spectrographs are well-suited to cover their large spatial extent. Spectroscopic exploration has been performed for a few GC peripheries \citep[e.g.,][]{wan_dynamics_2021,wan_dynamics_2022,2022MNRAS.512..315K,2025A&A...693A..69A}, and some kinematics signatures expected in N-body models, such as increased tangential anisotropy, have been uncovered. Though photometric explorations using chemically sensitive band-passes have been limited, they are key to identifying the faintest stars that populate the peripheries of GCs. Most recently, \citet[][]{2022mnras.515.5802B} identified multiple populations within the GD-1 using the photometric sensitivity of the CN lines in the LAMOST bands $u$, $g$ and $i$ \citep[][]{2012RAA....12..723Z}. This unique approach to understanding the stellar populations within tidal streams demonstrates the power of using multi-band photometry as an extra discriminatory tool. 

An extra dimension of metallicity can be provided by the Pristine survey \citep[][]{2017MNRAS.471.2587S}. The Pristine survey is a narrowband photometric program that targets the [Fe/H]-sensitive CaHK lines using the MegaCAM imager on the Canada-France-Hawaii Telescope. The first data release \citep[][]{2024A&A...692A.115M} provides photometric metallicities for millions of stars and is cross-matched with Gaia, which provides the precision astrometry. This combination is paying dividends with some major discoveries about the structure of the MW, including the discovery of the very metal-poor nature of the stellar stream C-19 \citep[][]{2022MNRAS.514.3532E}. Most recently, the combination of Pristine and Gaia has been used to search for extra-tidal stars in the peripheries of Omega Centauri. \citet[][hereafter KI25]{2025MNRAS.537.2752K} used the Pristine-Gaia-Synthetic catalogue, which provides synthetic CaHK photometry and metallicities for all stars in Gaia with BP/RP coefficients. KI25 identified stars in the tidal tails of Omega Centauri with a metallicity distribution similar to that observed within the GC, marking the first evidence of complex stellar populations from Omega Centauri being present among the tidally stripped stars. This demonstrates that Pristine provides a great opportunity to study the peripheries of GCs, assessing which stars populate any debris and how that compares to the progenitor GC.

This paper continues the analysis completed in KI25, taking the techniques developed therein and applying them to a set of GCs present within Pristine. The following section will introduce the Pristine data and target selection. After we will discuss the methods used to complete the Pristine data analysis, compare those results to GC orbits, known streams, and N-body and particle spray models, before providing our concluding comments.

\section{The Data}\label{sec:The_Data}
This work utilizes the Pristine-Gaia-Synthetic (PGS) catalogue from Pristine DR1 \citep[][]{2017MNRAS.471.2587S,2024A&A...692A.115M}. This catalogue provides all-sky synthetic, de-reddened photometry for sources from Gaia. Specifcally, it uses the Gaia $\mathrm{{G}_0}$, $\mathrm{{G}_{BP,0}}$, and $\mathrm{G_{RP,0}}$ information \citep[][]{2021A&A...649A...3R} to calculate synthetic CaHK-band photometry. Each source is associated with its photometric $\FeHPr$, as the CaHK-band can serve as a proxy for metallicity \citep[see][]{2017MNRAS.471.2587S}. PGS also provides a cross-matching between itself and Gaia, offering astrometry alongside the CaHK synthetic photometry. We searched all GCs in the PGS catalogue and visually inspected them to determine if they were well-sampled enough (i.e., enough stars within the tidal radial radius to identify the GC population to proceed, which typically corresponded to at least 100 stars. We found that 31 GCs met our visual inspection criteria; however, since NGC 5139 was the focus of KI25, we present the results for the remaining 30 GCs. These GCs are listed, along with some properties, in Table \ref{tab:GC_list}. To search for tidally stripped stars, we defined a radius of 5 degrees about the cluster center. This size allows for maximum detection of extra-tidal stars while minimising the negative effects that can arise when surveying a large section of the sky, such as projection effects and large variable extinction.

The PGS catalogue underwent a series of cuts to provide the cleanest data set for our purposes. We retained stars that met the following criteria:
\begin{itemize}
    \item $\mathrm{-4<[Fe/H]_{PGS}<0}$, the optimum range for Pristine photometric metallicities
    \item $\mathrm{\sigma_{{[Fe/H]}_{PGS}}<0.3}$ dex, to only use highly accurate measurements
    \item $\mathrm{P_{var}<0.3}$, to remove variable stars
    \item The corrected flux excess, $\mathrm{C^{*} < 3\sigma_{C^{*}}}$, and Renormalised Unit Weight Error $\mathrm{RUWE < 1.4}$, to remove stars with poor astrometric and photometric solutions \citep{2018A&A...616A...2L,2021A&A...649A...2L}
    \item Parallaxes that place a star beyond a heliocentric distance of 3 kpc, to remove all nearby stars.
\end{itemize}

The cleaned catalogues underwent a series of coordinate transformations that deal with projection effects in the large clustercentric radius explored here. The positions ($\alpha,\delta$) were transformed to a gnomonic coordinate system that deprojects the curvature in the plane of the sky to a flat $(x,y)$ coordinate plane centred on a GC whose coordinates are $(\alpha_{0},\delta_{0})$. The transformation follows the equations \citep[][]{2018A&A...616A..12G}:
\begin{equation}
\begin{aligned}
x = \cos\delta\sin(\alpha-\alpha_\mathrm{0})\\
y = \sin\delta\cos\delta_\mathrm{0}-\cos\delta\sin\delta_\mathrm{0}\cos(\alpha-\alpha_\mathrm{0}) \\
\end{aligned}
\label{eq:tangent_spat}
\end{equation}
\noindent
The proper motions ($\mu^{*}_\alpha,\mu_\delta$)\footnote{$\mu^{*}_\alpha\equiv\mu_{\alpha}\cos{\delta}$} also transformed to ($\mu_x,\mu_y$), through the following transformation:
\begin{equation}
\begin{aligned}
\mu_{x} = \mu_{\alpha}^{*}\cos(\alpha-\alpha_{\mathrm{0}})-\mu_{\delta}\sin\delta\sin(\alpha-\alpha_{\mathrm{0}})\\
\mu_{y} = \mu_{\alpha}^{*}\sin\delta_\mathrm{0}\sin(\alpha-\alpha_\mathrm{0})\\
 \quad+\mu_{\delta}\left(\cos\delta\cos\delta_\mathrm{0}+\sin\delta\sin\delta_\mathrm{0}\cos(\alpha-\alpha_\mathrm{0})\right)
\end{aligned}
\end{equation}
\noindent

and the uncertainties and covariances are propagated through. We corrected each star about each cluster centre for solar reflex motion in the proper motions using the defined solar motion of (12.8, 245.6, 7.78) km s$^{-1}$ \citep[][]{2018RNAAS...2..210D}, and assuming the heliocentric distance to each cluster from \citet[][]{baumgardt_accurate_2021}.

Despite inspecting each GC for the effects of variable reddening, the surveyed area of GCs will still be impacted by extinction. This may impact our ability to determine whether any detected overdensity is a true feature. Therefore, we followed the same technique presented in section 2.2 of \citet[][]{2020MNRAS.495.2222S}. Beginning at $G_0 = 17$ mag and increasing in brightness in steps of 0.05 magnitude, we retrieved all stars with a 0.25 magnitude range. We created a 2D density distribution which placed stars in 6 x 6 arcmin bins ranging between a clustercentric radius of 1.5 - 5 degrees. The log star counts in the distribution are fit with a first-order polynomial, and the root mean square (RMS) is calculated. The faintest $G_0$ magnitude with a guaranteed RMS less than 0.05 dex was regarded as the appropriate depth for exploration. In most cases, this was around 17th magnitude and brighter in some cases of regions of high variable extinction or high field density. 

After all these steps have been applied to the PGS data sets, we obtained a dataset that contains synthetic metallicities, transformed spatial and proper motion coordinates, and de-reddened photometry. Across each field of view explored, we have between $10^5$ and $10^6$ stars that are parsed through the methods in the next section to identify potential GC member stars.

\section{Methods}\label{sec:Methods}
\begin{figure*}
\figurenum{1}
\includegraphics[width=\textwidth]{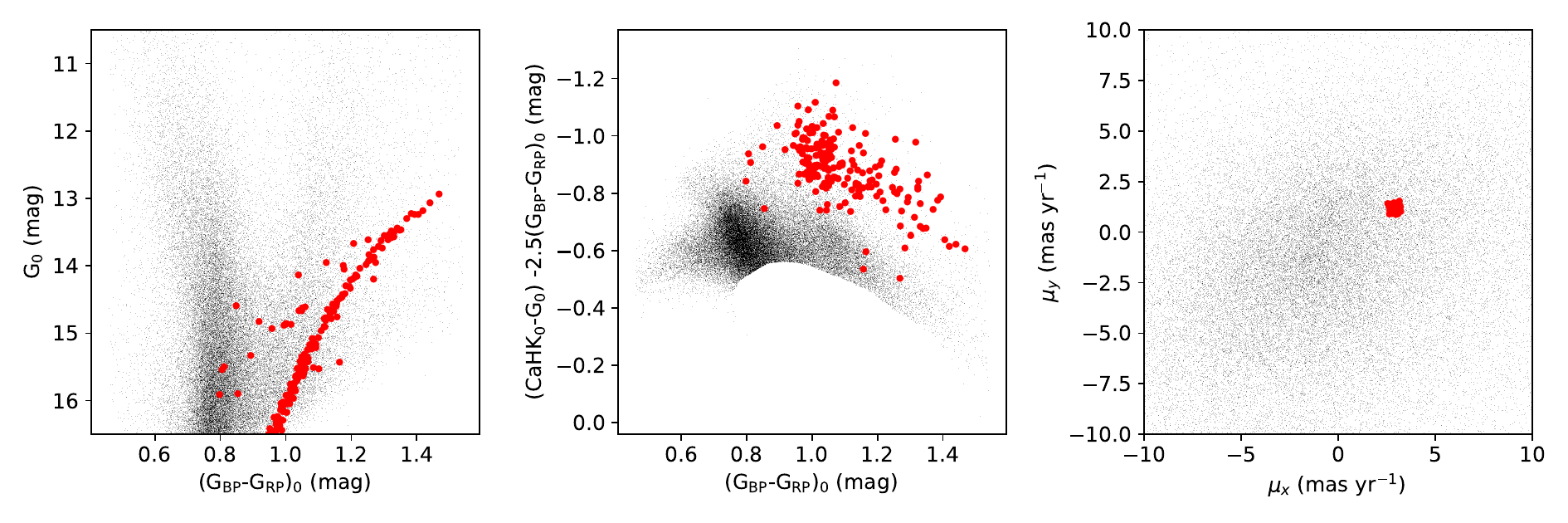}
\caption{Demonstration of the parameter spaces explored for the two reference sets with respect to NGC 7089. Points in red belong to the cluster sample, and the black points are the field sample. Left: CMD space. Center: The colour-colour space. Right: The proper motion space in tangential coordinates. An extended version of this figure for all GCs is located in the appendix.}\label{fig:CMD}
\end{figure*}

In KI25, we utilised unsupervised machine learning to identify GC-like stars in the periphery of Omega Centauri. We apply the same technique to the other 30 GCs in our sample. We will briefly describe the underlying techniques here, but refer the reader to KI25 for an in-depth presentation. We adopted a probabilistic model which identifies GC-like stars across three parameter spaces: proper motion space, colour-magnitude (CMD) space, and colour-colour (CC) space. The two former spaces utilise the astrometry ($\mu_x,\mu_y$) and photometry ($\mathrm{(G_{BP}-G_{RP})_{0}, G_0}$) from Gaia, while the latter space utilises the PGS CaHK-band, which is sensitive to [Fe/H]. Specifically, the relationship of the Gaia colours $\mathrm{(G_{BP}-G_{RP})_{0}}$ and the colour index $\mathrm{(CaHK - G_0) - 2.5(G_{BP} - G_{RP})_0}$ (hereafter $\mathrm{CaHK_{ind}}$) can be used to identify [Fe/H]-similar groups. We define two reference samples for each GC, representing how the field and GC populations are distributed across the three-parameter spaces. The GC reference sample was created by considering stars within each GCs tidal radius \citep[][]{de_boer_globular_2019}, whose proper motion is within 1 mas yr$^{-1}$ of the GC and 1.5 mas yr$^{-1}$ in the case of NGC 104. This creates a very clean GC sample, with no/very little field contamination (e.g., Fig.  \ref{fig:CMD}). As for the field sample, we utilised all stars outside of two degrees, maximising the field coverage, while minimising the number of GCs stars persisting in the reference sample. Some GCs in our sample have known extended structure, but the overall number of extra-tidal stars is assumed to be severely outnumbered by the field and, therefore, are of no concern in the field sample.


We model each component with their own likelihood function. CMD and CC space both follow a $k$-nearest neighbour function, where $k$ = 10th nearest neighbour. Given the difference in errors between colour and magnitude, and colour and colour, we adopted a weighting metric which represents the ratio between the typical uncertainties between Gaia colours and the Gaia G-band photometry, which is 1:10 in CMD space, and in $\mathrm{CaHK_{ind}}$ index in CC space the metric is 1:2. The likelihood function then takes the form of:
\begin{equation}
\ln P_{\rm{\rm{CMD/CC}}} = \ln\left(\frac{10}{\pi ((\frac{{\rm{\chi_{1}}}}{m_x})^2+(\frac{{\chi_{2}}}{m_{y}})^2)}\right)-\ln(N)  \label{eq:pcmd}
\end{equation}

where ($\chi_1$,$\chi_2$) is different in value between the star in question and its 10th nearest neighbour. In CMD space, those parameters are ($\chi_1$,$\chi_2$) = ($\mathrm{(G_{BP}-G_{RP})_0, G_{0}}$) and in CC space ($\chi_1$,$\chi_2$) =  ($\mathrm{(G_{BP}-G_{RP})_0, CaHK_{ind}}$), and $m_{x}$ and $m_{y}$ are the previously mentioned metric values. Additionally, $N$ is the size of the reference sample. The proper motion space adopts a 2D normal distribution centred on the bulk motion of the two reference samples. Given the proper motion vector ($\mu_x,\mu_y$), associated uncertainty ($\sigma_{\mu_x},\sigma_{\mu_y}$) and corresponding covariance, $p$, the 2D normal distribution takes the form of:

\begin{equation}
\ln{P_\mathrm{PM}}=-\frac{1}{2}(X-\bar{X})^\top\,C^{-1}\,(X-\bar{X}) - \frac{1}{2}\ln{4\pi\det[C]} 
\label{eq:lpm}
\end{equation}
\noindent

where $X = (\mu_x,\mu_y)$ is the proper motion of our targets, and $\bar{X}$ is the bulk motion of the cluster/field. $C$ is the covariance matrix, and it takes the form:
\begin{equation}
C=\begin{bmatrix}
\sigma^2_{\mu_{x}}+\sigma^2_{x}& p\sigma_{\mu_{x}} \sigma_{\mu_{y}}\\
p\sigma_{\mu_{x}} \sigma_{\mu_{y}}& \sigma^2_{\mu_{y}}+\sigma^2_{y}\\
\end{bmatrix} 
\label{eq:cov}
\end{equation}
\noindent
where $p$ is the associated correlation between the $\mu_{x}$ and $\mu_{y}$ directions, and ($\sigma_{x},\sigma_{y}$) is the dispersion of the field/GC sample proper motions in the $x$ and $y$ direction. We combine these probabilities to form a likelihood function to assign each star a membership probability that belongs to either the GC or the field. We can utilise these previous functions to identify a value $f_\mathrm{cl}$, the ratio of cluster-to-field stars in our sample, providing normalisation in the likelihood function. That likelihood function takes the form of:

\begin{equation}
L = f_{\mathrm{cl}} P^{\mathrm{cl}}_{\mathrm{PM}}P^{\mathrm{cl}}_{\mathrm{CMD}}P^{\mathrm{cl}}_{\mathrm{CC}} + (1-f_{\mathrm{cl}})P^{\mathrm{MW}}_{\mathrm{PM}}P^{\mathrm{MW}}_{\mathrm{CMD}}P^{\mathrm{MW}}_{\mathrm{CC}}
\end{equation}

where $P^{\mathrm{cl/MW}}$ is the probability function as discussed previously concerning either the GC sample, denoted by $\mathrm{cl}$, or the field sample, denoted by $MW$. We assigned each star a membership probability ($\mathrm{P_{mem}}$), between 0 and 1, where 1 is a likely GC member and 0 is a possible field contaminant, by the following equation:

\begin{equation}
\mathrm{P_{mem}} = f_{\mathrm{cl}} P^{\mathrm{cl}}_{\mathrm{PM}}P^{\mathrm{cl}}_{\mathrm{CMD}}P^{\mathrm{cl}}_{\mathrm{CC}}/L
\end{equation}

This $\mathrm{P_{mem}}$ value is calculated for each star in each 5-degree field explored. Once the final step above is complete, we aim to visualise the 2D surface density distribution of GC stars. For each GC, we placed a grid between -5 and 5 degrees in both directions with knots at every 6 arcmin. At each knot in the grid, we calculated the local density of GC-likely stars:

\begin{equation}
\rho(x,y) = \frac{N_{xy}}{\pi R^2_{\mathrm{loc}}}
\end{equation}

where $N_{xy}$ is the number of nearest stars to the knot that satisfy the sum of the corresponding $\mathrm{P_{mem}}$ is at least 2, and $\mathrm{R_{loc}}$ is the largest radius of the stars from the knot when the above condition is met. This technique applies variable smoothing to the two-dimensional density profile. In areas where there are many high-probability stars, the smoothing, or \( R_{\mathrm{loc}} \), will be minimal. Conversely, in regions lacking high-probability stars, the smoothing will be greater. Additionally, we consider the effects of any MW field density gradient across any field of view by fitting a 2D first-order polynomial in ($x, y$) to the above 2D surface density profile while excluding the inner 2 degrees to avoid the influence of the GCs in the fitting. The resulting polynomial is subtracted from the 2D surface density profile, removing those field gradients. Lastly, we calculated the average knot density ($\overline{\rho}$) and associated standard deviations ($\sigma_\rho$) across the field view, again excluding the overwhelm counts of the inner two degrees of the field of view to avoid the GC central regions from biasing the statistics, and used these values to present each knot in the 2D density distribution as the number of standard deviations about the mean value: $(\rho-\overline{\rho})/\sigma_\rho$.

We aimed to explore both the \fehpgs~radial distribution and the 2D surface density distributions of the GCs in our sample. However, after completing the analysis, we noted that some GCs in our sample had incomplete Gaia scanning coverage or lacked enough extra-tidal stars to produce a \fehpgs~radial distribution or a 2D surface density map. Eight GCs, namely NGC 288, NGC 2298, NGC 4590, NGC 5466, NGC 5897, NGC 5904, NGC 6681 and NGC 6809, we exclude from the upcoming discussion based on these issues. We present our findings on these GCs in the appendix \ref{sec:GC_removed}.

\section{Results}
\subsection{Extra-tidal Candidates and Metallicity Distributions}\label{sec:metal_dist}
The techniques described in the previous section generate probabilities indicating the likelihood that a star belongs to either the GC or the MW field. Based on the calculated probabilities, we determined the \fehpgs~ values for the GCs. First, considering stars with the GCs tidal radius, similar in definition to the GC reference samples, we calculated the weighted \fehpgs~ mean, utilising the measurement uncertainties and \Pmem as weights, and present those results in Table \ref{tab:feh_list}, along with the number of stars used in performing this calculation. We find, much like the results presented in \citet[][]{2024A&A...692A.115M}, fantastic agreement between the photometric metallicities of Pristine and the known spectroscopic metallicities of \citet[][]{2009AA...508..695C}, with mean difference between the two measurements as $0.0005 \pm 0.03$.

At increasing radial distance, we venture into the regions where little is known about the metallicity distribution of the GC populations. In Fig. \ref{fig:FeH_part1}, we demonstrate the radial metallicity distribution for each GC, and for ease of interpretation, we split the stars into different groups and present them separately. For stars with $\mathrm{P_{mem}>0.5}$, we demonstrate their membership probability by the colour of the marker. As for stars $\mathrm{P_{mem}} < 0.1$, which provides the underlying field distribution, we present them as a density distribution. Within each GCs tidal radius (inner dotted vertical line), we see the strong signatures of the GC, and in most cases, the presence of stellar population between the tidal and Jacobi (outer dashed vertical line), and several high probability stars into the GC periphery. In some cases where there are GCs within the target GC field of view, such as in the case of NGC 104 and NGC 362, and NGC 6218 and NGC 6254 (this GC is not included in our sample), there is a clear separation between them. The interloping GCs do not contribute to the target GCs radial metallicity distribution, as their stars have been correctly identified as non-members.

One of the aims of this work is to detect extra-tidal stars and explore the metallicities of those stars, comparing them to the GC we assume they originate from. To do this, we calculated two different weighted mean metallicity values for each GC. In addition to the GC measurement previously discussed, we explored the metallicity between the tidal radius and the Jacobi radius ([Fe/H]$^{\mathrm{r}_{\mathrm{t}}}_{\mathrm{PGS}}$), and outside the Jacobi radius ([Fe/H]$^{\mathrm{r}_{\mathrm{J}}}_{\mathrm{PGS}}$). These radial ranges correspond respectively to the outer regions of bound stars where the potential escapers are distributed and nearing stripping, and to stars that have already been stripped. Exploring these ranges can infer that the same population, on average, is consistent throughout each radial range or if there are notable differences in the populations. However, before measuring the metallicity of the outer regions, we must address the possibility of contamination in the regions beyond the Jacobi radius. Across our sample, in Fig.  \ref{fig:FeH_part1}, there are signs of potential contamination even at modest probability levels (\Pmem $>$ 0.5). In the metallicity region of \fehpgs$>-1.0$ dex, where the field population is dominant, there is an increase in metal-rich stars as we approach \Pmem = 0.5 (e.g., NGC 6205 or NGC 6218). These potential contaminants can shift the weighted mean metallicities in the metal-rich direction. To mitigate this, we defined a conservative sample for each GC (see also section 4.2 in KI25), which only includes stars of $\mathrm{P_{mem}}>0.99$, and this creates the cleanest selection of high-probability stars for analysing the outer regions. In Fig.  \ref{fig:FeH_part1}, the conservative sample is identified as circled points, and across all GCs there is a clear consistency with the conservative sample and those within the tidal radius, and are a sure platform to explore the mean metallicities of the GC outer regions.

The weighted mean of the two radial regions is calculated in the same method as the GC sample. In this case, we only used stars in the conservative sample, and weights were constructed from the measurement uncertainties and \Pmem. Table \ref{tab:feh_list}, presents the mean metallicities across the three ranges. Fig.  \ref{fig:FeH_part1} illustrates the mean values in relation to the radial metallicity distribution for each GC. Additionally, Fig.  \ref{fig:feh_rad_compare} allows us to compare the differences among the three mean measurements. In most cases, there is consistency across the radial regions covered, and there is a predominantly clear one-to-one relationship between the mean measurements across the three radial regions, which supports our assumption that the stars are excited/stripped from the GC central regions. We acknowledge that our conservative selection approach may directly influence our findings, as we are focusing on the stars most likely to be members of the GC. However, this selection method allows us to obtain the most accurate estimate of the mean metallicity for populations located beyond the tidal and Jacobi radii. Repeating the above analysis towards \Pmem = 0.5, we found that the metallicities increased across the board, typically on the order of 0.2 dex, indicating the presence of contamination. We assess the contamination level at \Pmem = 0.5 by counting the number of stars outside the Jacobi radius whose \fehpgs was consistent with [Fe/H]$^{\mathrm{r}_{\mathrm{J}}}_{\mathrm{PGS}}$, and taking the ratio of that number to all stars present. The ratio across our sample has an average value of 20 per cent, which decreases to 15 per cent when we consider only GCs with more than 10 stars with \Pmem$>$0.5 dex, indicating that high contamination only exists amongst GCs with low numbers of high probability members. Considering the potential contamination at even modest levels of membership probability, more discriminatory tools are required before establishing more stars as bona fide members at decreasing membership probability. We provide the membership probabilities for all stars analysed as online material\footnote{10.5281/zenodo.15832846}.

\begin{deluxetable*}{lcrcrcrr}
\tablecaption{The mean $\mathrm{[Fe/H]_{PGS}}$ across three radial ranges in the conservative samples; within the tidal radius ([Fe/H]]$^{\mathrm{cl}}_{\mathrm{PGS}}$), between the tidal radius and Jacobi radius ([Fe/H]$^{\mathrm{r}_{\mathrm{t}}}_{\mathrm{PGS}}$), and outside the Jacobi radius ([Fe/H]]$^{\mathrm{r}_{\mathrm{J}}}_{\mathrm{PGS}}$). Their 1$\sigma$ uncertainty accompanies each measurement, and the number of stars in the conservative sample, $N_C$, in the calculation. Regions with one star have that star's PGS measurement presented instead. Additionally, we include the number of stars with $\mathrm{P_{mem}}>0.5$, $N^{r_j}_{>0.5}$, that are consistent with the metallicities of the conservative sample. Regions with zero stars have no value presented. \label{tab:feh_list}}
\tablewidth{0pt}
\tablehead{
\colhead{Cluster} &
\colhead{[Fe/H]$^{\mathrm{cl}}_{\mathrm{PGS}}$} &
\colhead{$N^{cl}_{C}$} &
\colhead{[Fe/H]$^{\mathrm{r}_{\mathrm{t}}}_{\mathrm{PGS}}$} &
\colhead{$N^{r_t}_{C}$} &
\colhead{[Fe/H]$^{\mathrm{r}_{\mathrm{J}}}_{\mathrm{PGS}}$} &
\colhead{$N^{r_j}_{C}$}&
\colhead{$N^{r_j}_{>0.5}$}}
\startdata
NGC 104 & $-0.71\pm0.12$ & 1548 & $-0.64\pm0.18$ & 20 & $-0.67\pm0.33$ & 43& 140 \\
NGC 362 & $-1.23\pm0.23$ & 195 & $-1.30\pm0.26$ & 8 & $-1.28\pm0.42$ & 6& 23\\
NGC 1261 & $-1.35\pm0.06$ & 25 & $-1.26\pm0.38$ & 4 & $-1.48\pm0.09$ & 3& 7\\
NGC 1851 & $-1.27\pm0.27$ & 78 & $-1.18\pm0.42$ & 17 & $-1.56\pm0.21$ & 10& 19\\
NGC 1904 & $-1.52\pm0.06$ & 70 & $-1.45\pm0.17$ & 2 & $-1.59\pm0.22$ & 13& 24\\
NGC 2808 & $-1.18\pm0.19$ & 67 & $-1.34\pm0.64$ & 4 & $-1.43\pm0.38$ & 8& 27\\
NGC 3201 & $-1.58\pm0.22$ & 389 & $-1.38\pm0.18$ & 33 & $-1.43\pm0.13$ & 3& 4\\
NGC 5272 & $-1.55\pm0.15$ & 449 & $-1.34\pm0.22$ & 1 & $-1.65\pm0.16$ & 2& 4\\
NGC 6101 & $-2.09\pm0.10$ & 79 & $-2.00\pm0.20$ & 12 & $-2.19\pm0.19$ & 1& 2\\
NGC 6205 & $-1.54\pm0.07$ & 366 & $-1.39\pm0.09$ & 1 & $-1.49\pm0.24$ & 10& 35\\
NGC 6218 & $-1.38\pm0.08$ & 159 & $-1.47\pm0.17$ & 1 & $-1.32\pm0.09$ & 7& 27\\
NGC 6341 & $-2.36\pm0.09$ & 137 & $-2.35\pm0.11$ & 9 & $-2.5\pm0.26$ & 2& 2\\
NGC 6362 & $-1.10\pm0.23$ & 283 & $-1.12\pm0.30$ & 4 & $-1.04\pm0.34$ & 13& 52\\
NGC 6397 & $-2.23\pm0.19$ & 447 & $-2.23\pm0.11$ & 43 & $-2.15\pm0.13$ & 2& 4\\
NGC 6541 & $-1.62\pm0.10$ & 85 & $-1.56\pm0.08$ & 2 & $-1.57\pm0.20$ & 3& 47\\
NGC 6584 & $-1.49\pm0.37$ & 19 & $--$ & 0 & $-1.36\pm0.69$ & 7& 42\\
NGC 6752 & $-1.51\pm0.10$ & 550 & $--$ & 0 & $-1.46\pm0.12$ & 14& 62\\
NGC 6934 & $-1.39\pm0.04$ & 26 & $-1.44\pm0.13$ & 1 & $-0.95\pm0.43$ & 5& 36\\
NGC 6981 & $-1.39\pm0.02$ & 32 & $--$ & 0 & $-1.33\pm0.22$ & 5& 11\\
NGC 7078 & $-2.28\pm0.09$ & 197 & $--$ & 0 & $-1.73\pm0.21$ & 5& 11\\
NGC 7089 & $-1.47\pm0.10$ & 147 & $-1.5\pm0.18$ & 6 & $-1.70\pm0.29$ & 4& 11\\
NGC 7099 & $-2.26\pm0.02$ & 123 & $--$ & 0 & $-2.27\pm0.06$ & 2& 4\\
\enddata
\end{deluxetable*}

\begin{figure*}
\figurenum{2}
    \centering
    \includegraphics[width=1.9\columnwidth]{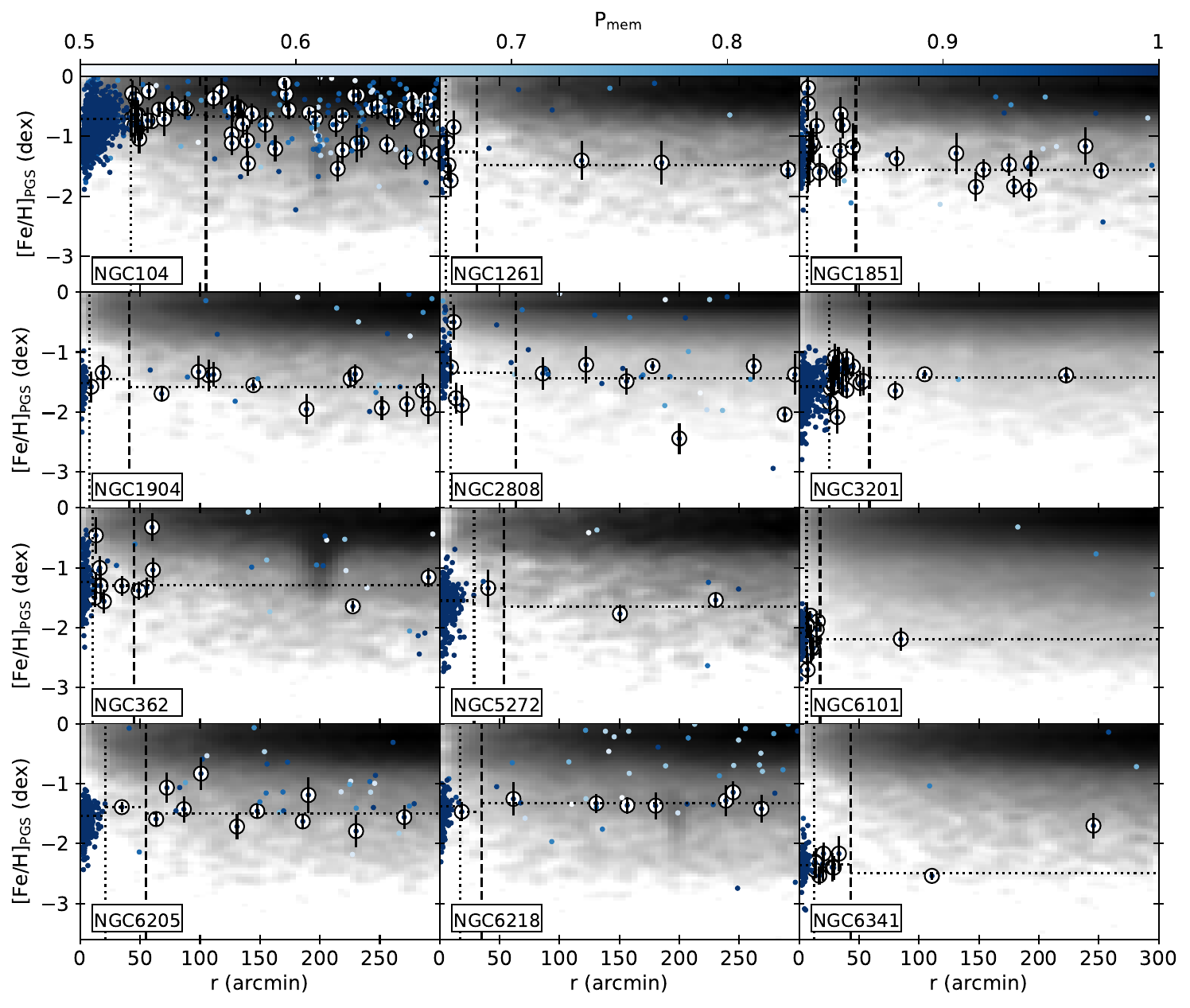}
    \caption{Radial distribution of $\mathrm{[Fe/H]_{PGS}}$ for our sampled GCs. Stars with $\mathrm{P_{mem}>0.5}$ are displayed and coloured by their membership probability (top colour bar). Underplotted is a density diagram that demonstrates the distribution of stars with $\mathrm{P_{mem}<0.1}$ (most likely field members). Each star that is part of the conservative sample for each GC is marked with a black circle, along with their $\mathrm{[Fe/H]_{PGS}}$ uncertainty displayed with error bars. The vertical dotted (dashed) lines indicate the tidal (Jacobi) radius, and the black horizontal lines show the mean $\mathrm{[Fe/H]_{PGS}}$ in the conservative sample.}
    \label{fig:FeH_part1}
\end{figure*}

\begin{figure*}
\figurenum{2}
    \centering
    \ContinuedFloat  
\includegraphics[width=1.9\columnwidth]{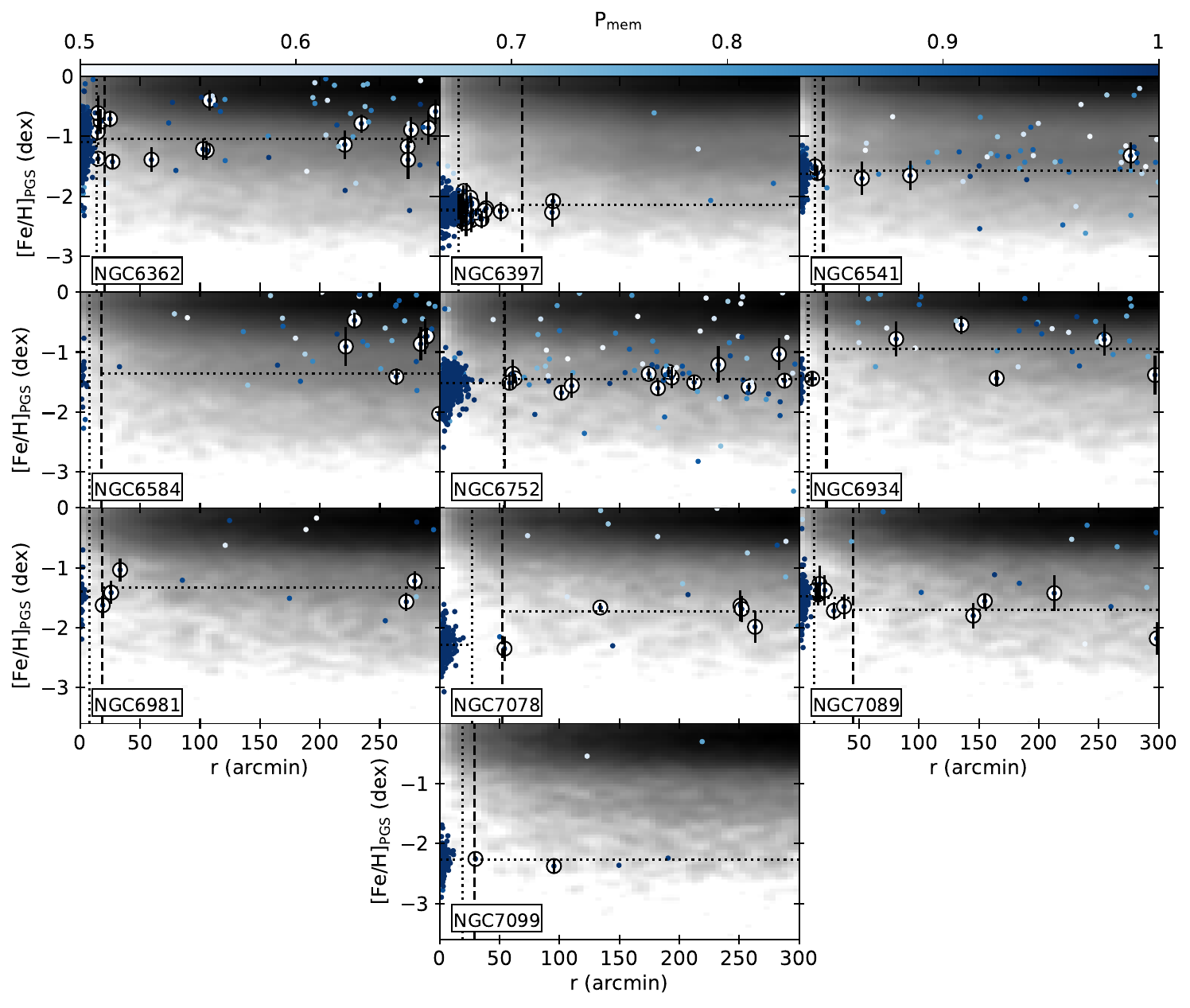}
    \caption[]{Radial distribution of $\mathrm{[Fe/H]_{PGS}}$ for our sampled GCs. A continuation of Fig. \ref{fig:FeH_part1}.}
    \label{fig:FeH_part2}
\end{figure*}

\begin{figure*}
\figurenum{3}
    \centering
    \includegraphics[width=\textwidth]{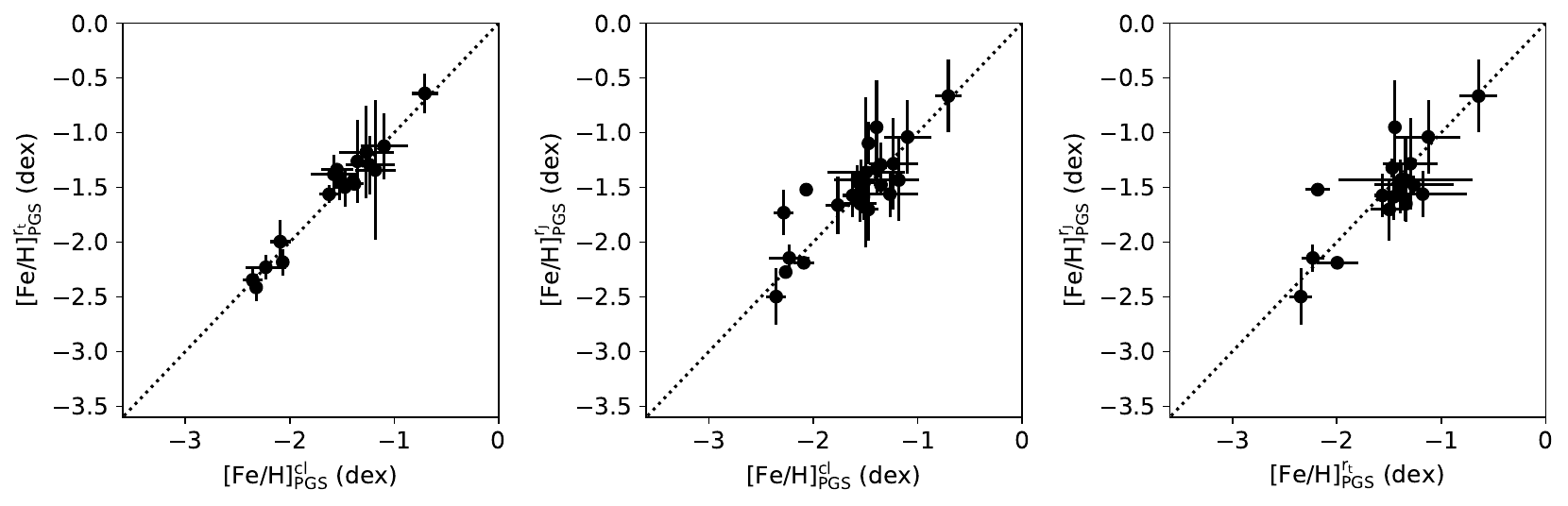}
    \caption[]{Comparison of the mean $\mathrm{[Fe/H]_{PGS}}$ measurements across the three radial ranges for each GC. The radial ranges explored are deonted by the x- and y- axis labels, where $\mathrm{[Fe/H]^{r_{cl}}_{PGS}}$ denotes the radial range that covers within the tidal radius, $\mathrm{[Fe/H]^{r_t}_{PGS}}$ between the tidal and Jacobi radius, and  $\mathrm{[Fe/H]^{r_J}_{PGS}}$ outside the Jacobi radius. A one-to-one correspondence is indicated by the dashed diagonal line.}
    \label{fig:feh_rad_compare}
\end{figure*}

\subsection{2D Surface Density Distributions}
 For 22 GCs, we created the 2D distribution following the methods described in section \ref{sec:Methods}, and present them in Fig.  \ref{fig:1_part1}. Along with the 2D distribution, we include any known stream in the vicinity of each GC that is present in the \texttt{Galstreams} catalogue \citep[][]{10.1093/mnras/stad321}, and include the orbit integrated with a \texttt{MWPotential2014} from \citet[][]{2014ApJ...795...95B}. Our results can trace the directions of known extended structure across multiple GCs according to \texttt{Galstreams}. Those GCs are NGC 1851, NGC 3201, NGC 6341, NGC 6362, NGC 6397, NGC 7089 and NGC 7099. Intriguingly, there is a disagreement between the \texttt{Galstreams} tracks and the observed debris surrounding NGC 1261, and there are several reasons this could be (e.g, projection effects, photometric depth, see section \ref{sec:disagreement}). Several GCs with extended structure not present in \texttt{Galstreams} are also showing similar signs of elongation and extended debris, such as NGC 362 \citep[][]{2019MNRAS.486.1667C}, NGC 1904 \citep[][]{2020MNRAS.495.2222S}, NGC 2808 \citep[][]{2020MNRAS.495.2222S}, NGC 5272 \citep[][]{2023ApJ...953..130Y} and NGC 6981 \citep[][]{2021A&A...646A.176P}. We tentatively find newly identified extended structure around six GCs: NGC 6101, NGC 6205, NGC 6218, NGC 6541, NGC 6752, and NGC 6934.

 \subsubsection{Classification of Extended Structure}
Given the variety of shapes around all the GCs studied here, we aimed to classify the debris as either an asymmetric structure (tidal tail-like) or spherical/irregular in shape (diffuse envelope-like). Therefore, we calculated the position angles of the identified extended structure as a function of radius. For each 2D map, starting from the GC center and increasing in clustercentric radius, we sorted each knot into its corresponding radial bin and calculated the position angle of that bin, beginning at north and rotating eastward. Then, we fit a squared cosine function similar to the quadrupole spatial density model of \citet[][]{2021MNRAS.507.1127K} (see Fig. 5 therein) to the densities as a function of position angle. The corresponding shift of the squared cosine peak is recorded as the position angle of the GC in that radial bin. This application achieves two goals. First, the measurement as a function of radius assesses the amount of isophotal twisting seen in the debris. Second, we can take a global position angle measurement of the debris by finding the mean position angle of the debris outside the Jacobi radius, where we expect extended structure to begin. This is done by calculating the weighted mean ($\overline{\theta}$) and associated uncertainty $\sigma_{\overline{\theta}}$ for the position angle measurements based outside the Jacobi radius, and out to the extent of the 2 sigma contours seen in Fig. \ref{fig:1_part1}. The uncertainty $\sigma_{\overline{\theta}}$ can be used to asses the axisymmetric nature of the extended structure. That is, if the uncertainty on the position angle for a given structure is suitably small, the position angle is highly constrained. This is expected in the case of clear axisymmetric structure akin to tidal tails. The opposite would be true in the case of any irregular/spherical structure akin to diffuse stellar envelopes. In Fig.  \ref{fig:sig_Posa}, we show $\sigma_{\overline{\theta}}$ as a function of heliocentric distance, excluding NGC 104, as there is no continuous extended structure outside the Jacobi radius of the GC (see Fig. \ref{fig:1_part1}), and display the position angle as function of clustercentric radii in Fig \ref{fig:PosA_part1}. Inspecting Fig. \ref{fig:sig_Posa}, there is a clear dichotomy between those with tightly constrained position angle measurements and those who are loosely constrained outside the Jacobi radius. We separated the sample into two groups: the tidal tail group, where $\sigma_{\overline{\theta}}<8$ deg, and the envelope group, where $\sigma_{\overline{\theta}}>8$ deg. We therefore present 17 GCs that have tidal tail extended structure, and 4 GCs have a diffuse envelope/irregular structure. Out of the new tentative detections, we find NGC 6101, and NGC 6541 are classified as possessing diffuse envelope structure, and NGC 6205, NGC 6218, NGC 6752 and NGC 6934 most likely possess structure in the form of tidal tails.


\subsubsection*{Comparison to Models}
Beyond the detections of PGS, it is worth comparing the debris to expectations from N-body models. Conducting our own is considered beyond the scope of this work. Instead, we compare our surface density distributions to the N-body models of the \textsc{eTidalsGCs}\footnote{https://etidal-project.obspm.fr/} project \citep[hereafter eTidals,][]{2023AA...673A..44F}. Briefly, the models of eTidals take the present-day MW GCs structural and kinematic parameters, integrate their orbit backwards by 5 Gyr, and then let a N=100000 particle GC evolve forward to the present day, with a Plummer model representing the GC, and a series of different potentials for the MW halo. For our purposes, we accessed the publicly available PII models of eTidals, which include a thick and thin disk, and a spherical dark matter halo, but no bulge component. We transformed the spatial coordinates of eTidals to the same gnomonic tangential coordinate system as the Pristine data. Lastly, we created the 2D density distributions, presented in Fig. \ref{fig:1_part1_etidals}, and the position angle profile in the same way. The resulting position angle profiles as a function of radius are presented in Fig.  \ref{fig:PosA_part1}, as with the position angles of the orbit as a function of radius in the forward and backward directions and the direction of the Galactic Centre. Across 22 GCs, there are varying levels of agreement with the direction of the Pristine density distributions and eTidals, or the position angles of the orbit or the Galactic Centre. This is a useful comparison for GCs that don't show clear extended structure, such as NGC 104, which does not show strong signs of extended structure, but its position angle matches the eTidals model. Several GCs show firm agreement across the whole radial range explored with eTidals (e.g., NGC 3201, NGC 6218, NGC 6362, NGC 7099), some show partial agreements, either rotating away from the eTidals model (e.g., NGC 6981) or towards it (e.g., NGC 6101). We also see some GCs pointing along the orbit while not directly aligning with eTidals (e.g., NGC 7089).  Given the variety of extended structure, and the different levels of accuracy when compared to expectations of N-body models and our understanding of GC disruption, constructing a complete picture of the ubiquity of extended structure in MW GCs may be a difficult task, but large homogeneous surveys of GCs are a great place to start.

\begin{figure*}
\figurenum{4}
    \centering
    \includegraphics[width=1.8\columnwidth]{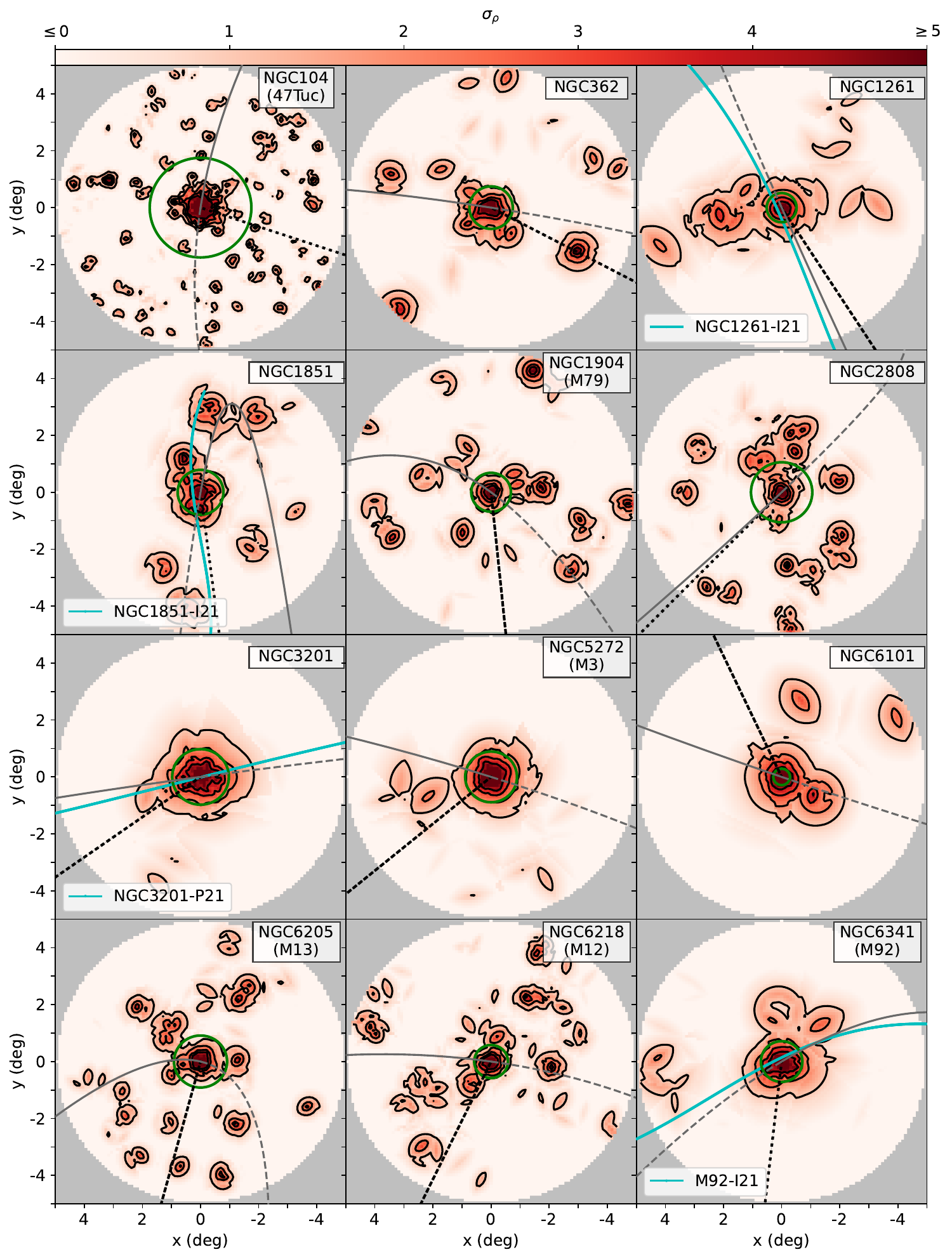}
    \caption{2D density distribution of our analysed GCs. Each GC, labelled in the top right, contains their surface density distribution coloured by $\sigma$ level on the top row. The contours indicate 1, 2 and 3 $\sigma$ levels. Additionally plotted are the GCs orbit (forward/backward orbits are indicated by the solid/dashed grey line) based on a \texttt{MWpotential2014} \citep[][]{2014ApJ...795...95B}, the direction towards the Galactic Center (dotted black line), the GCs Jacobi radius (green circle), and the \texttt{Galstreams} stream track of any extended structure (cyan line). \textit{Stream track labels: I21 - \citet[][]{2021ApJ...914..123I}, P21 - \citet[][]{2021MNRAS.504.2727P}.}}
    \label{fig:1_part1}
\end{figure*}

\begin{figure*}
\figurenum{4}
    \centering
    \ContinuedFloat
    \includegraphics[width=1.8\columnwidth]{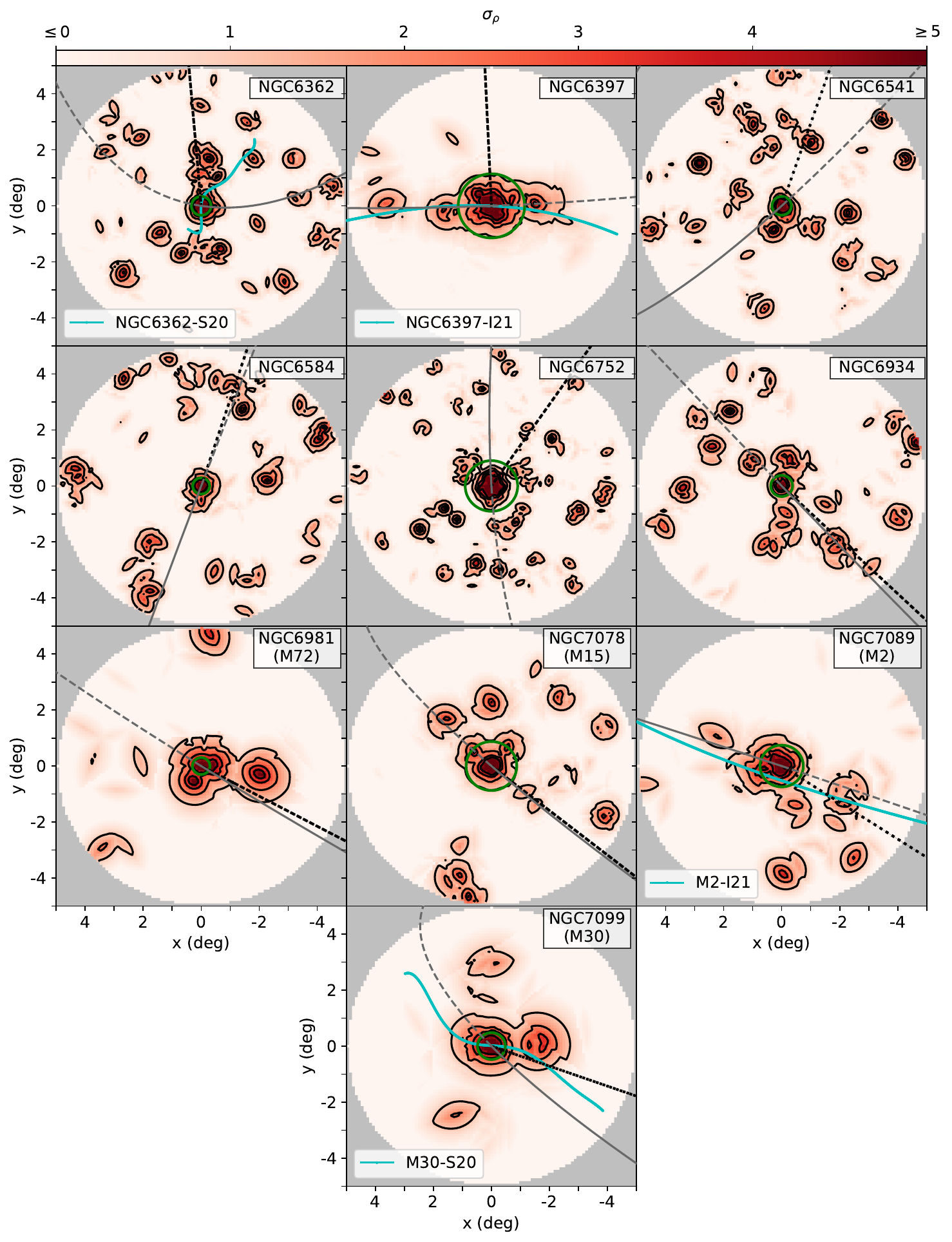}
    \caption[]{2D Density distribution of our analysis GCs. A continuation of Fig.  \ref{fig:1_part1}. \textit{Stream track labels: S20 - \citet[][]{2020MNRAS.495.2222S}, I21 - \citet[][]{2021ApJ...914..123I}.} }
    \label{fig:1_part2}
\end{figure*}

\begin{figure}
\figurenum{5}
    \centering
    \includegraphics[width=\columnwidth]{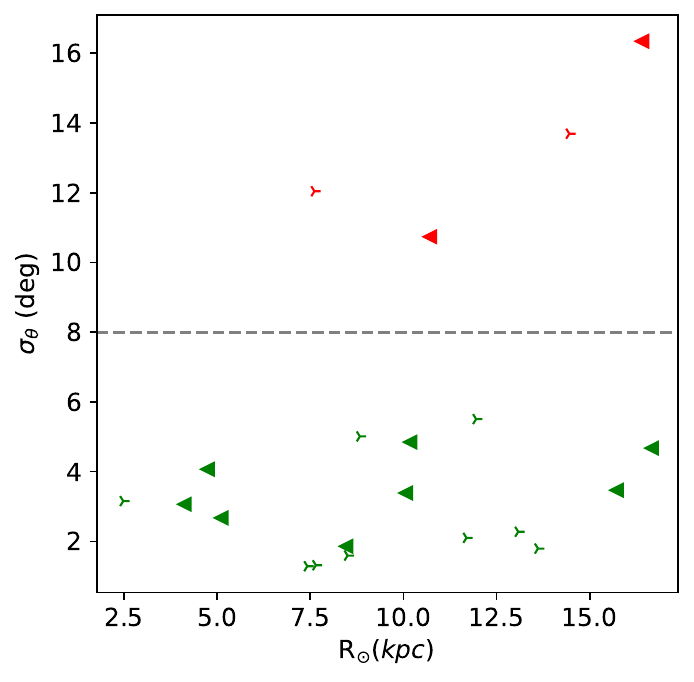}
    \caption[]{Demonstration of the separation between extended tidal structure that is highly constrained (green), and those that are less constrained (red), designated by $\sigma_{\theta}$ as a function of heliocentric distance, $\mathrm{R_{\odot}}$. The pointing direction of the arrows (triangles) indicated whether a GC last passed the periocenter (apocenter) of their orbit. The separation of $\sigma_{\theta}=8$ is shown by the horizontal dashed line.}
    \label{fig:sig_Posa}
\end{figure}

\begin{figure*}
\figurenum{6}
    \centering
    \includegraphics[width=1.8\columnwidth]{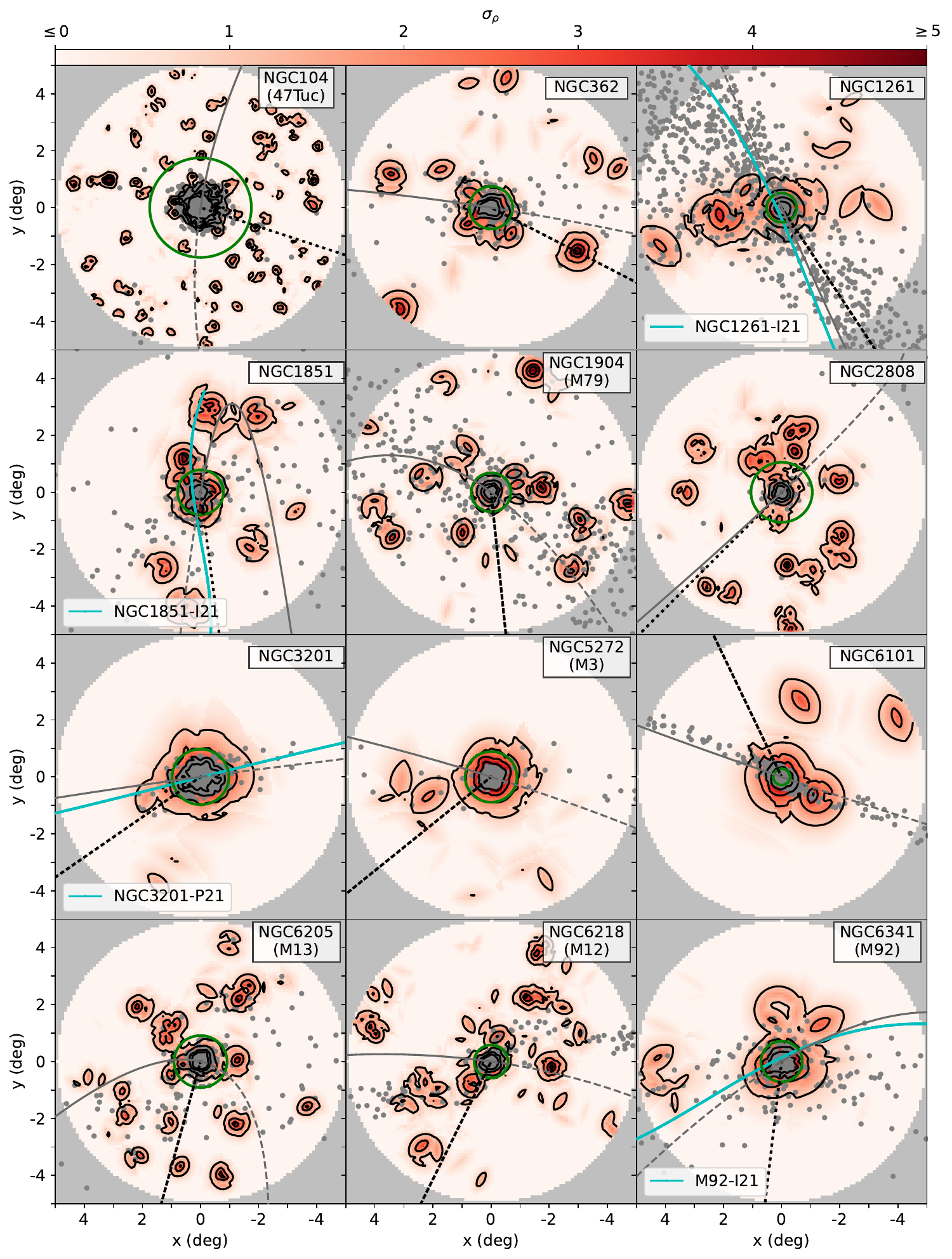}
    \caption{The same as Fig.  \ref{fig:1_part1}, with the addition of the eTidals N-body simulations of disrupting GCs, demonstrated with the grey points. \textit{Stream track labels: I21 - \citet[][]{2021ApJ...914..123I}, P21 - \citet[P21][]{2021MNRAS.504.2727P}.}}
    \label{fig:1_part1_etidals}
\end{figure*}

\begin{figure*}
\figurenum{6}
    \centering
    \ContinuedFloat
    \includegraphics[width=1.8\columnwidth]{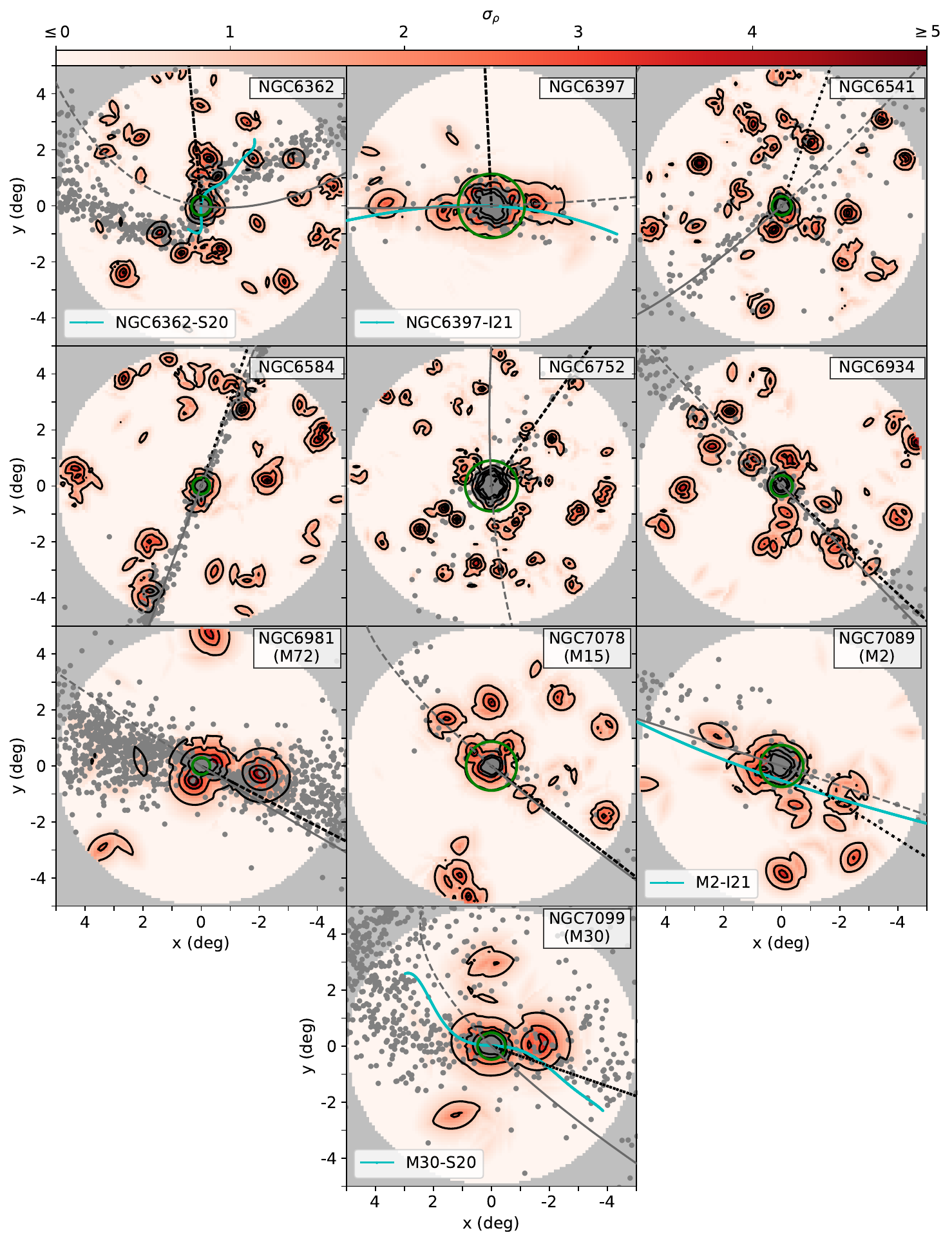}
    \caption[]{A continuation of Fig.  \ref{fig:1_part1_etidals}. \textit{Stream track labels: S20 - \citet[][]{2020MNRAS.495.2222S}, I21 - \citet[][]{2021ApJ...914..123I}.}}
    \label{fig:1_part2_etidals}
\end{figure*}

\begin{figure*}
\figurenum{7}
    \centering
    \includegraphics[width=\textwidth]{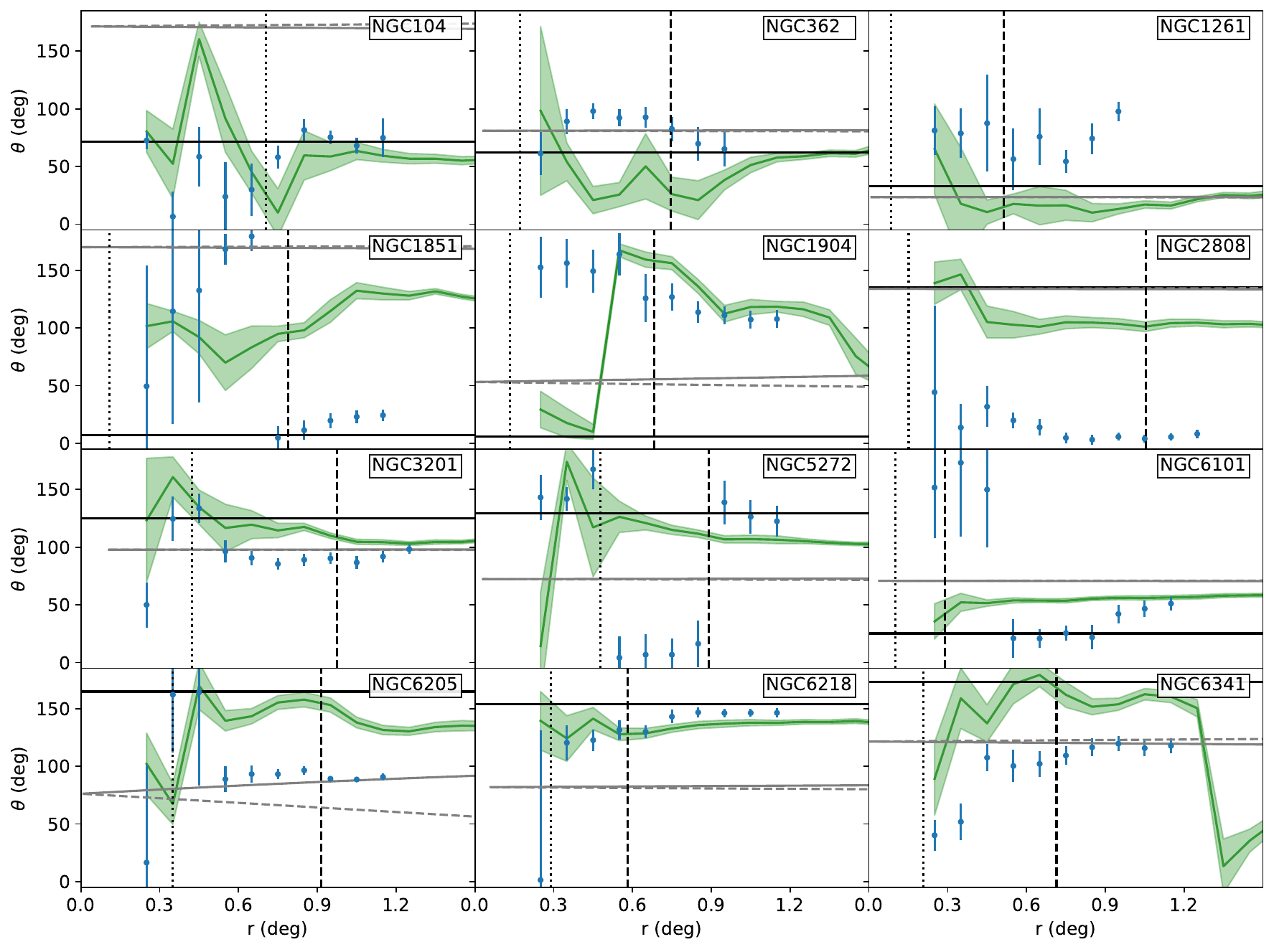}
    \caption{Position angle as a function of radius for our GCs with 2D surface density distributions. Each GC, labelled in the top right corner, has its radial position angle profile demonstrated by the blue points. Also demonstrated are the forward/backward orbits as solid/dashed grey horizontal lines, and the position angle of the eTidal N-body models is demonstrated by the solid green line, and the shaded region shows the 1$\sigma$ uncertainties of the measurements. Last, the GC tidal and Jacobi radii are shown through dotted and dashed vertical lines, respectively.}
    \label{fig:PosA_part1}
\end{figure*}

\begin{figure*}
\figurenum{7}
    \centering
    \ContinuedFloat
    \includegraphics[width=\textwidth]{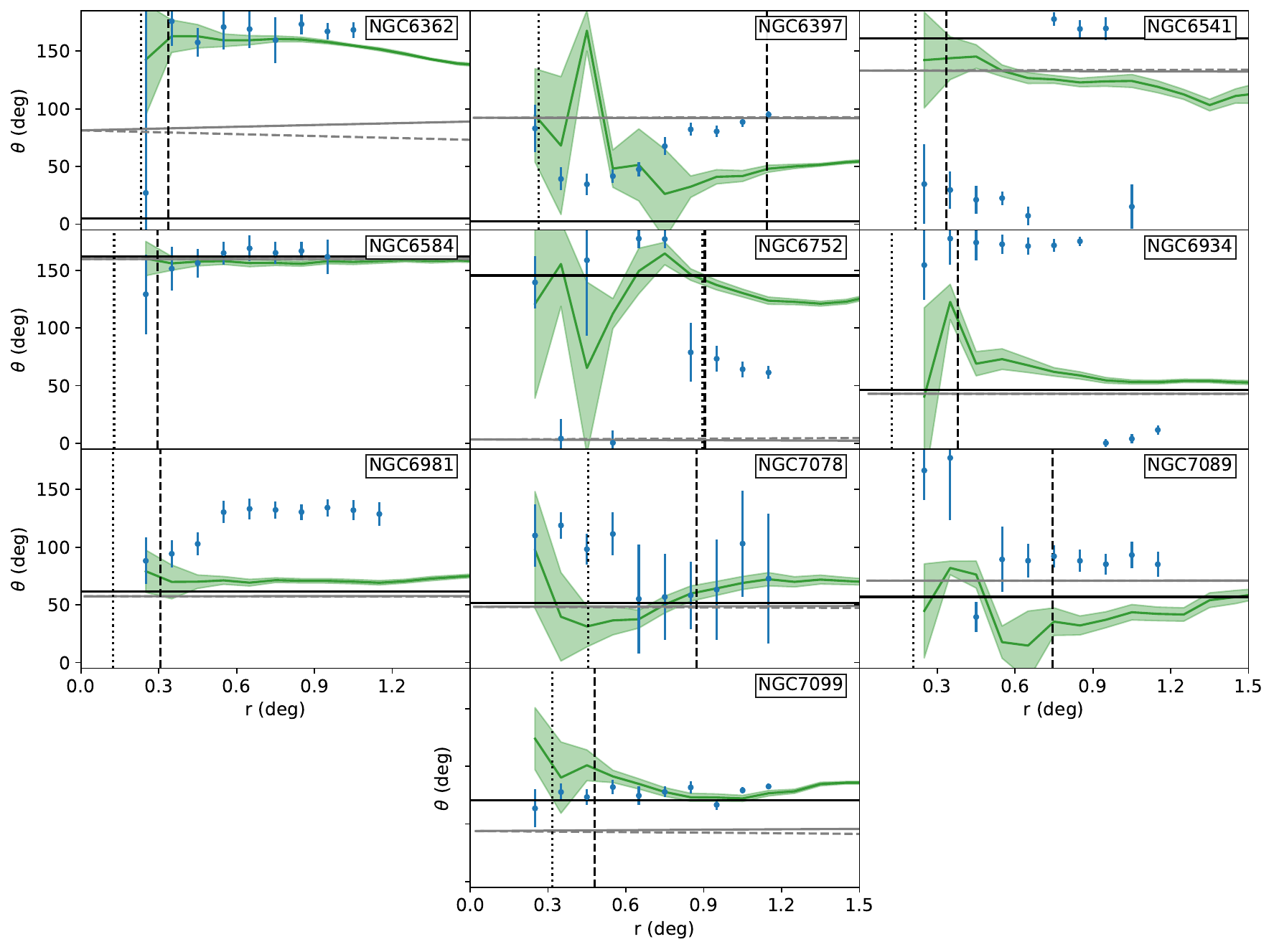}
    \caption[]{Position angle as a function of radius for our GCs with 2D surface density distributions. A continuation of Fig.  \ref{fig:PosA_part1}.}
    \label{fig:PosA_part2}
\end{figure*}

\begin{figure*}
\figurenum{8}
    \centering
    \includegraphics[width=\textwidth]{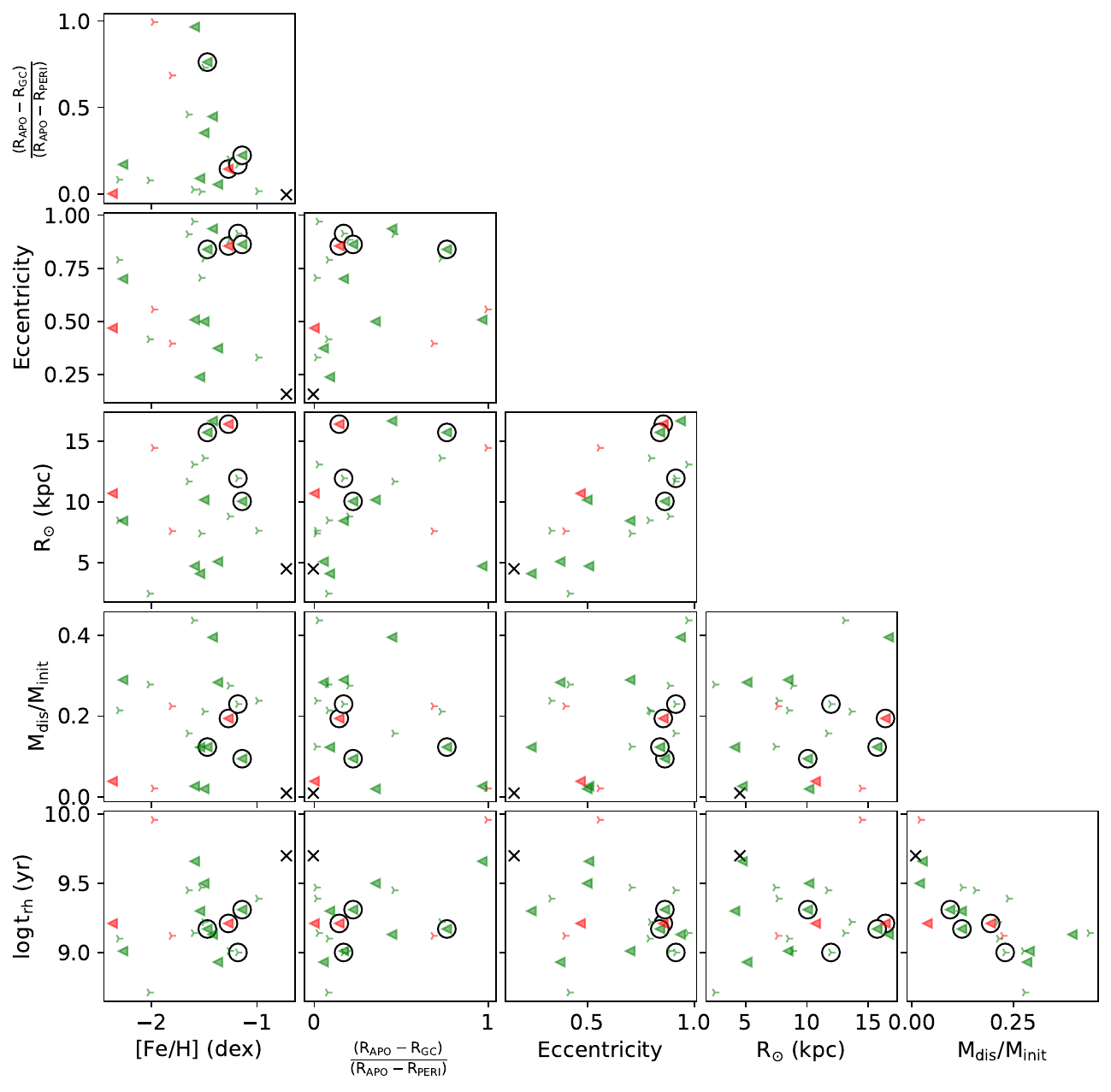}
    \caption[]{Corner plot of GC parameters used to explore potential predictors of extended structure and its form. Points share colours and triangle orientations with Fig.  \ref{fig:sig_Posa}. NGC 104 is marked with a cross; NGC 1261, NGC 1851, NGC 2808, and NGC 6934 are circled in black. Parameters from top to bottom: orbital phase, eccentricity, heliocentric distance, mass loss fraction, and relaxation time; left to right: \fehpgs, orbital phase, eccentricity, heliocentric distance, and mass loss fraction.}
    \label{fig:param_spaces}
\end{figure*}

\section{Discussion}

\subsection{Assessment of Parameter Spaces for Detected Structure}
Our exploration of PGS offers a chance to explore whether there is any parameter space (structural or kinematics) that may be used to identify whether a GC may possess extra tidal features. We may posit that GCs on more eccentric orbits and/or smaller apo/pericenters may experience more tidal forces/disk passages, which can lead to the development of extended tidal structure than those on more circular orbits or those that rarely approach the disk of the MW. Further, GCs that have a shorter relaxation time (the time for the motions of the stars within the GC to become completely randomised) may have undergone mass loss, which will naturally speed up their relaxation time \citep[][]{2019ApJ...882...98P}. A study compiled by \citet[][]{2020A&A...637L...2P} demonstrated that there is no discernible parameter space that may hint at a GC possessing extended structure.  Here, we provide our take on this problem with a set of homogeneous detections within one dataset. We also considered the proposed origins of these GCs as presented in \citet[][]{massari_origin_2019} and \citet[][]{2025arXiv250314657M}.

Our attempts to classify any parameter space as a predictive tool for GCs with tidal debris paint a complicated picture. Much like the analysis before by \citet[][]{2020A&A...637L...2P}, we do not find any significant grouping that could be used to identify GCs which may possess extended tidal structure. In Fig.  \ref{fig:param_spaces}, we demonstrate a selection of parameter spaces that may best be used to separate GCs:
\begin{itemize}
    \item Orbital phase of the GC in the form of $(\mathrm{R}_{\mathrm{apo}}-\mathrm{R}_{\mathrm{GC}})/(\mathrm{R}_{\mathrm{apo}}-\mathrm{R}_{\mathrm{peri}})$
    \item Orbital eccentricity defined as $\mathrm{(R_{apo}-R_{peri})/(R_{apo}+R_{peri})}$ \citep[][]{2019ApJ...882...98P}
    \item Heliocentric distance \citep[$\mathrm{R}_{\odot}$,][]{baumgardt_accurate_2021}
    \item Metallicity \citep[\FeH,][]{2009AA...508..695C}
    \item The fraction of mass lost as defined by $\mathrm{M_{dis}/M_{tot}}$, following the calculations of \citet[][]{2020A&A...637L...2P} and  
    \citet[][]{2018MNRAS.478.1520B}
    \item the relaxation time \citep[$\log{\mathrm{t}_{\mathrm{rh}}}$,][]{2019MNRAS.485.1460S}
\end{itemize}

Apart from expected relationships such as orbit eccentricity and heliocentric distance, these selected parameter spaces overall show no clear relationship between themselves, nor with our classifications of tidal tails or irregular/diffuse spherical structure.  Further, we do not see any distinct correlation between whether a GC is approaching peri/apocentre, nor whether a GC is tagged to any specific accretion event or formed {\it in-situ}. It is worth noting here that, given the photometric depth that we are probing within PGS, we are biased to large, nearby GCs, whereas \citet[][]{2020A&A...637L...2P} completed their analysis compiling all the known GCs extended structure from literature, which includes GCs beyond the limiting R$_{\odot} \approx 16$ kpc in our work. A loose connection in our data is that amongst the most metal-poor GCs ($\mathrm{[Fe/H]_{PGS}}$ $< -1.8$ dex), there are as many GCs with envelope structure as those that show asymmetric features. However, further analysis with deeper photometry is required to assess whether this is a true relationship.

\subsection*{Assessment of Detected Extended Structure}
Over recent decades, many studies have explored the disruption process in GCs. The general understanding is that when stars in GCs are excited into higher orbits, they can be stripped from the GC by passing through the Lagrange points, where the gravitational forces of the MW and host GC are in balance \citep[e.g.,][]{kupper_tidal_2010,kupper_more_2012}. Stars stripped in this manner will typically follow the orbit of their progenitor GC, creating the long iconic axisymmetrical tidal tails. However, the Lagrange points generally point towards the Galactic centre, so there are points along the orbit of a GC where the direction of stripped stars does not align with the orbit or any tidal stream. This creates isophotal twisting in the extended structure, which is demonstrated visually by the characteristic s-shape that is prevalent in N-body models of tidal streams \citep[e.g. eTidals,][]{2023AA...673A..44F}. Taken together, these points show that the shape of extended tidal structures around GCs can vary for each GC, depending on its orbital phase or even the shape of its orbit. According to the N-body models of \citet[][]{2015MNRAS.446.3100H}, a GC nearer to Apogalacticon may be pointing towards the Galactic center, with the overall stream track following the orbit. Once a GC is traversing between apo- and perigalacticon, stars ejected may be quite close to the orbit \citep[see Fig.  2 in][]{2015MNRAS.446.3100H}. In our case, we observe these effects to varying degrees. For example, we find 10 GCs are currently within two kpc of their apocenter, five of which show tidal tails pointing in the direction of the Galactic center, and three other structures pointing in the direction along the orbit. Furthermore, some GCs exhibit obvious isophotal twisting, identified by changes in position angle as a function of radius (e.g., NGC 3201 and NGC 6205). Encouragingly, the morphologies of the GCs in PGS correspond well with the eTidals simulations. This also comments that on the surface, these halo models utilised in eTidals are useful for comparisons across all MW GCs.

\begin{figure*}
\figurenum{9}
    \centering
        \includegraphics[width=1.6\columnwidth]{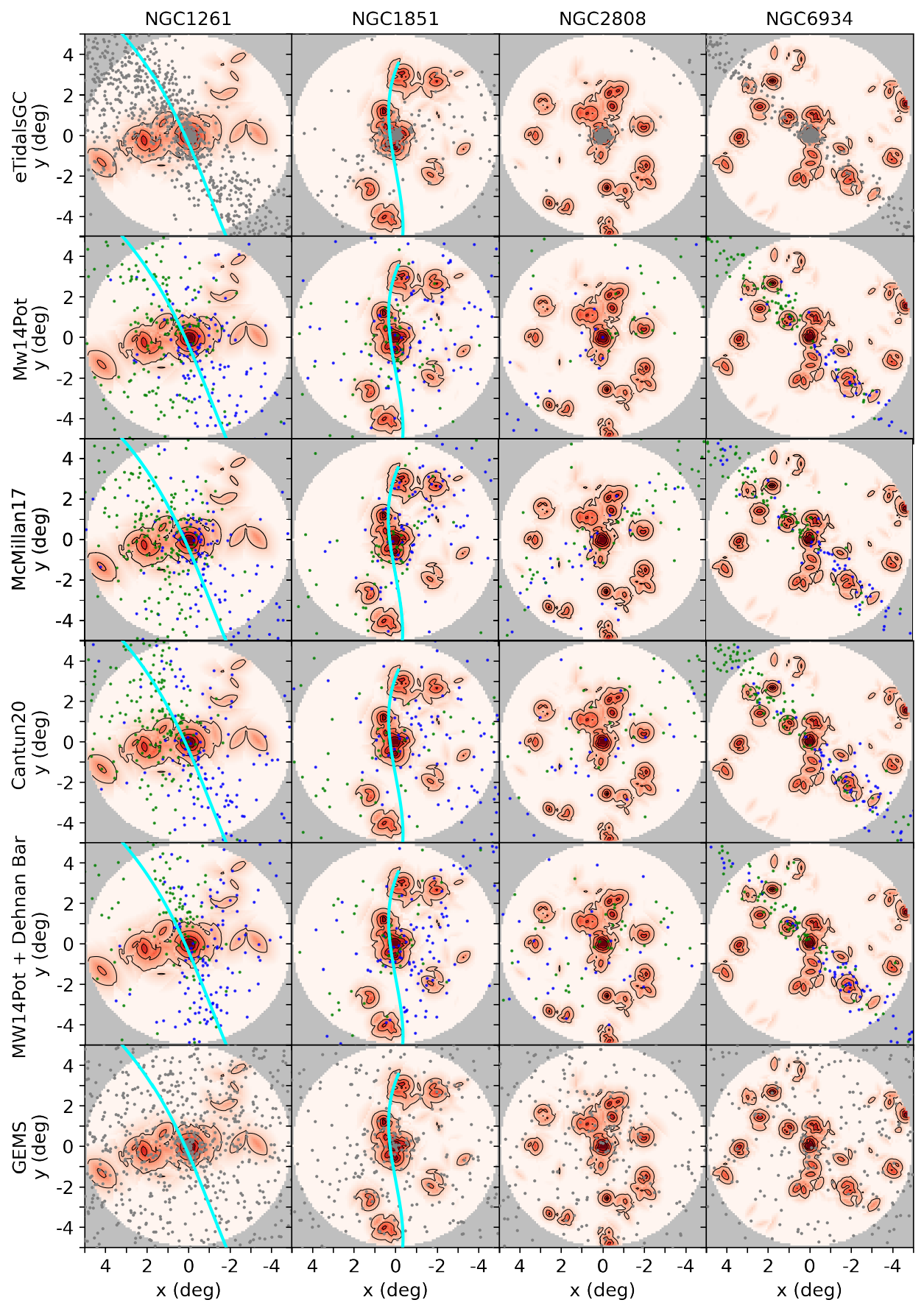}
    \caption[]{ Demonstration of the eTidalGCs N-body models (top row) the \citet[][]{2025ApJS..276...32C} particle spray models in both forward and backward trajectories in the under four selected MW potentials: \texttt{McMillan17} \citep[][]{2017MNRAS.465...76M}, \texttt{Cautan20} \citep[][]{2020MNRAS.494.4291C}, and a \texttt{MWpotential2014} \citep[][]{2014ApJ...795...95B} both with and without a \citet[][]{1999ApJ...524L..35D} bar, and randomly selected N=10000 particles from GEMS \citep[][]{2024MNRAS.528.5189G} (bottom row). The forward trajectory, with leading tail points marked in green, and the backwards trajectory, with trailing points marked in blue. They are overplotted on the 2D density profiles of the four GCs NGC 1261, NGC 1851, NGC 2808 and NGC 6934 (left to right columns), and the \texttt{Galstreams} stream track is also presented. The density profiles, contours and \texttt{Galstreams} track are the same as what is presented in Fig.  \ref{fig:1_part1}.}
    \label{fig:spray_compare}
\end{figure*}

\subsection{Disagreement between Observations and Theory}\label{sec:disagreement}
However, it is worth noting that there are cases of disagreement between eTidals and disruption expectations, the \texttt{Galstreams} stream tracks, the orbital path and some GCs. The egregious cases are NGC 1261, NGC 1851, NGC 2808 and NGC 6934. There are several reasons why this could be the case. The first is the shape of the Galactic potential used to model the disruption of the GCs. N-body models such as eTidals and \citet[][]{2015MNRAS.446.3100H} are dependent on many moving parts. This includes the shape of the GC halo, disk and dark mass. Not only that, but it is also dependent on the initial conditions of the progenitor system that is left to evolve. This includes structural parameters such as mass and the kinematics that underlie the orbit, which are essential for the disruption modelling process. If the models of GC disruption do not match the data, it may indicate that any of these components are incorrect. Since the four GCs have relatively eccentric orbits and are close to their apocenter (except one), we consider them sensitive to the MW potential. Another factor in any discrepancy is the overall method for tidal tail generation. Models like eTidals simplify interactions within the GC as it is computationally expensive to consider N-body interactions within the GC core during long integrations/orbital passages. The Globular cluster Extra-tidal Mock Stars catalogue \citep[hereafter GEMS,][]{2024MNRAS.528.5189G}, explores the shape of the distribution of extra-tidal stars of GCs, considering only three-body interactions within the core of the GCs, providing a new perspective on the formation of the disruption process of GCs. 

Considering these discussed effects, we utilised both the particle spray model techniques of \citet[][]{2025ApJS..276...32C} and the publicly available GEMS database\footnote{https://zenodo.org/records/8436703}\citep[][]{2024MNRAS.528.5189G}. The particle spray models are completed in both the forward and backward directions of the orbit, and within four selected MW potentials: \texttt{McMillan17} \citep[][]{2017MNRAS.465...76M}, \texttt{Cautan20}\citep[]{2020MNRAS.494.4291C}, and a \texttt{MWpotential2014} \citep[][]{2015ApJS..216...29B} both with and without a \citet[][]{1999ApJ...524L..35D} bar. We adopted the orbital kinematics from \citet[][]{2021MNRAS.505.5978V}, which is based on Gaia DR3 measurements, and like eTidals, simulated the particle spray model across 5 Gyrs. While the accuracy of the potentials used in this analysis is beyond the scope of this work, it is still useful to assess how different the tidal debris can appear for each GC in varied potentials. As for the GEMs database, GEMS provides N=50000 extra tidal stars per GC, evolved over 5 orbits within a \texttt{MWpotential2014}.  In Fig.  \ref{fig:spray_compare}, we present the particle spray models and a randomly sampled N=10000 particles from the GEMS distribution with the 2D surface density profiles of the four GCs. There are immediate similarities between the debris across the GCs; however, subtle changes can affect how compact/broad the debris may appear on the plane of the sky, and on which GC is examined, implying that a global assessment is not a trivial task. However, we break down each of the four GCs to comprehend our observations.

The existence of extra-tidal structure around NGC 1261 has been known for the past decade. The first detection of extended structure was in the form of a tidal tail \citep[e.g.,][]{2000A&A...359..907L}, before evidence of a stellar envelope was presented with deep photometry \citep[e.g.,][]{2018MNRAS.473.2881K}. With the release of Gaia, \citet[][]{2024ApJ...967...89I} detected long tidal tails in the vicinity of this GC. The direction of this stream appears perpendicular to the direction of our findings in PGS. One possible interpretation of the shape of the extended structure is that, as NGC 1261 has recently passed apogalacticon, we are seeing the most recent debris lost in our 2D density distribution, while the underlying \texttt{Galstreams} line indicates the overall track of NGC 1261, wrapped over the field of view. This is similar to the shape of multiple mass-loss episodes across multiple passages presented in \citet[][]{2015MNRAS.446.3100H}. The evidence for stars perpendicular to the known stream track is also present in spectroscopic studies. \citet[][]{2025A&A...693A..69A} identified stars in this region as bona fide cluster members, suggesting that multiple passages and the existence of a bar can affect the debris shape, making the GC and its debris appear a lot broader on the sky. Comparing the particle spray models here, we can see how the different potentials make the tidal stream appear very broad, much like eTidals. The effect of a bar in the potential can completely change the direction of the tidal stream. Therefore, while we find debris inconsistent with the known stream track, our findings are not inconsistent with particle spray models and current observational identification of extra-tidal members.

NGC 1851 was first detected with an extended stellar envelope by \citet[][]{2009AJ....138.1570O}. Other deep photometric surveys also uncovered such a feature \citet[][]{2018MNRAS.473.2881K}, although tidal tails have been detected by \citet[][]{2018MNRAS.474..683C} pre-Gaia, and \citet[][]{2024ApJ...967...89I} post-Gaia. Our surface density profile is consistent with the direction of the known track, which traverses north to south, and is loosely consistent with models of eTidals at large radii, which point north-west to south-east. The particle spray models also agree mostly with the eTidal models, although a couple of models suggest the direction of the debris is closer to the known track. Much like NGC 1261, our findings are consistent with the current understanding of the direction of the known stream track.

\citet[][]{2018MNRAS.474..683C} uncovered a tail of NGC 2808 leading to the north-west with their deep, wide-field photometric study. The eTidals debris also travels in the same direction as the tail displayed in \citet[][]{2018MNRAS.474..683C}, but it is fairly low density in the field of view we are exploring. Our findings show strong signs of debris roughly perpendicular to the orbit/particle spray models. While the particle spray models follow the eTidals, implying limited dependence on the potential's shape, the addition of the bar introduces irregularities in the debris morphology. Our detected extended structure does appear to twist towards the direction of the orbit/eTidals stream at a large clustercentric radius. This may be evidence for a larger extended structure, and deeper photometry would confirm the existence of a long tidal stream.

NGC 6934 has not been reported to possess extended tidal structure. The eTidals representation of the NGC 6934 tidal tails shows clear, long, and thin tails traversing north-east to south-west. This direction is also mirrored in the particle spray models. However, our surface density profile of NGC 6934 shows north-to-south extensions, completely different to the compared models. However, at increasing clustercentric radii, there are several disconnected overdensities which can be traced through our conservative sample (see Fig.  \ref{fig:1_part2_cons}), therefore we may again be detecting the surface of a deeper tidal stream in the vicinity of this GC.

Across all four GCs, however, the GEMS catalogue provides a completely different view. GEMS modelling predicts that extra-tidal stars are found across the entirety of the field of view explored here, which is in complete contrast to the particle spray models and eTidals. This kind of distribution is a result of the random nature of ejected stars from the GC core, not through the Lagrange points, which is one of the main factors for the stream-like structure seen in models like in eTidals. What fraction of GC stars lost through this method with comparison to the more extensively covered methods like eTidals is unknown, but it is clear that the true distribution of tidal tails is most likely a combination of these effects.

The varying levels of agreement between PGS detections and different particle spray or N-body models highlight the fundamental challenges in understanding GC disruption across different gravitational potentials. Even the particle spray models, eTidals and GEMS exhibit both marginal and substantial differences in debris morphology across just these four cases—a small subset of the Milky Way GC population.  This underscores that the choice of potential can significantly influence the form of extended structures. Moreover, GC-specific parameters such as the orbital kinematics and structural parameters and associated assumptions, such as taking modern day GC properties and placing them 5 Gyrs back in time on their orbit, and then evolving them forward and simulating disruption, can lead to concerns about the accuracy of the debris formed and for comparisons to their modern day counterparts. The type of model of disruption is also worth noting. The difference in extra-tidal star distributions between eTidals and GEMS is stark, and shows how important it is to consider as many aspects of GC dynamical evolution as possible. Despite the different approaches these two works take, in truth, the disruption process of GCs would be some combination between low-mass stars excited into the peripheries and subsequently lost, and stars ejected from the GC through three-body relaxations, and it is worth considering different methods of dissolution to explain the tension between observations and models. Other matters that also concern the shape of extended structure include the effect of dark matter subhalos either embedding the host GCs \citep[e.g.,][]{penarrubia_stellar_2017}, or interacting with extended structure making it more disperse \citep[][]{2002MNRAS.332..915I} or chaotic orbits affecting stream density \citep[][]{2016MNRAS.455.1079P,2023ApJ...954..215Y}, all show that many pieces of the puzzle need to be considered to fully understand GC disruption and the shape extended structure may take. However, in this instance, there may be a simpler reason why there may be such disagreements between observations and theory.

\subsection{A case for disagreement: Photometric Depth}

A more likely scenario is the photometric depth of PGS. This body of work only scratches the surface of the 2D density distributions as PGS reaches G = 17 mag at best, which covers only the RGB in almost all cases (Fig.  \ref{fig:CMD}). Studies conducted using only Gaia \citep[e.g.,][]{2020MNRAS.495.2222S} utilise the full photometric range of Gaia, reaching G = 20 mag, and deep wide-field photometry such as DECam \citep[e.g.,][]{2024A&A...683A.151P} or Subaru \citep[e.g.,][]{10.1093/mnras/stab2325} can go considerably further still. The stars that are preferentially excited into the peripheries of GCs and subsequently are stripped are of lower mass, mainly main-sequence stars. The limits of PGS render these types of stars undetectable, hence, any extended structure populated by these types of stars will be missed \citep[see also][]{2018MNRAS.474.2479B}. This will make GCs in PGS appear not as elongated or tidal tails appear not as long, or not at all. Detecting more stars would improve contrast in the GC peripheries, which PGS lacks in this instance. However, as previously discussed, the fact that GCs in our sample are showing position angles that are consistent with the many models we have available to compare to, it is still useful to search datasets with faint photometric limits for elongation in the periphery, as this can indicate whether a GC is currently undergoing disruption, or even where searches can commence to identify those stars that stripped. 

Another effect of shallow photometry is that tidal tails may appear as overdensities in the field of view, disconnected from the GC body. In the 2D density distributions, there are many overdensities at varying significance levels. Given the typical rarity of high-mass stars in tidal tails, these structures may be true substructures indicative of underlying extended structure beyond the photometric limit of PGS. To demonstrate the validity of the detection of the substructures, Fig.  \ref{fig:1_part1_cons}) shows that there is generally a good alignment between the orbit and the conservative selections. The number of conservative stars in overdensities off-orbit or off-track may be the product of multiple episodes of mass loss as discussed previously. However, regions could just as easily be residual fluctuations in the MW field around the GCs observed. As seen in Fig.  \ref{fig:FeH_part1}, even at modest probability levels, we may still have significant numbers of field stars masquerading as GC stars (e.g., NGC 104). These field stars would have consistent proper motions, location on the CMD, and \fehpgs as the target GCs, and our technique would make those stars indistinguishable from GCs. The lack of detailed chemical information and, perhaps most critically, line-of-sight radial velocities limits our ability to accurately identify GC stars in their peripheries. This limitation is not unique to Pristine, as Gaia provides line-of-sight velocities only for relatively bright stars (G $\leq$ 14 mag). However, the imminent commencement of large-scale multi-object spectroscopic surveys such as WEAVE \citep[][]{Dalton2012WEAVE:Telescope,2024MNRAS.530.2688J} and 4MOST \citep[][]{2019Msngr.175....3D} can target stars as faint as Gaia to good accuracy, and the Prime Focus Spectrograph \citep[][]{2014PASJ...66R...1T,2016SPIE.9908E..1MT} can go even further still. These instruments will provide complete coverage of GCs and their peripheries and will uncover new and exciting extended structures, we look forward to what these surveys uncover.

\section{Conclusions}
The photometric survey Pristine has had its first data release, and we utilise it to search for extended structures around 30 MW GCs. Pristine's unique data set provides photometric metallicities, and this work takes advantage of the Pristine-Gaia-Synthetic catalogue, which contains all-sky synthetic metallicities cross-matched with Gaia sources. Combining the synthetic photometry and metallicities from Pristine and the Gaia photometry and proper motions, we used unsupervised machine learning, specifically k-nearest neighbours, to identify GC stars within a clustercentric radius of five degrees. Across 30 GCs, we identify high-probability stars outside the tidal radius. We measure the mean metallicities of GCs across three radial regions: within the tidal radius, between the tidal and Jacobi radii, and beyond the Jacobi radius. To minimise contamination from residual MW field stars, we constructed a conservative GC star sample with $P_{\mathrm{mem}} > 0.99$, using it to obtain the cleanest $\mathrm{[Fe/H]_{PGS}}$ estimates in the radial regions. Our results show consistency across all radial regions explored, suggesting that the highest-probability stars are tidally stripped.

We created 2D surface density distributions for the GCs to observe the distributions of the likely GC member stars. Through an adaptive smoothing technique, we provided the Pristine view of the known extended GC structure of 22 GCs, of which six are new tentative detections. Using the N-body models of eTidals and orbital information, we quantify the position angle of the GCs as a function of radius and compare our findings to the direction of GCs orbit, the eTidals N-body simulation and the expectations of GC disruption and dissolution. We find that Pristine performs admirably at identifying tidally stripped stars in the MW field. However, Pristine does have its weakness in its photometric depth. This lack of depth prevents us from claiming new detections as bona fide extended structure, but it does provide a strong basis for using CaHK-band or [Fe/H]-sensitive photometric bands to search for extended tidal structure. Further, our findings here, as amongst the largest homogenous survey of GC peripheries to date, demonstrate the ubiquity of extended structure amongst MW GCs.

As technology advances, our understanding of GC peripheries will continue to deepen. Deep, wide-field photometry will reveal new structures, while complementary large-scale multi-object spectroscopy will provide crucial missing information, such as line-of-sight velocities and detailed chemical tagging, not captured by current surveys like Gaia. Armed with all these tools, we will be able to completely separate the MW field from the comparatively less common GC stars in the periphery and start to fully understand how GCs evolve and dissolve and place them in the context of their contribution to the build-up of the Milky Way halo.

\begin{acknowledgments}
We thank the anonymous referee for the helpful comments which improved the quality of this paper. We wish to also thank Nicolas Martin and Julio Carballo-Bello for helpful discussions during the development of this work.
PBK acknowledges support from the Japan Society for the Promotion of Science under the programme Postdoctoral Fellowships for Research in Japan (Standard). This work was supported by JSPS KAKENHI Grant Numbers JP23KF0290, JP22K14076, JP21H04499, and JP20H05855. 

This work has made use of data from the European Space Agency (ESA) mission Gaia (https://www.cosmos.esa.int/gaia), processed by the Gaia Data Processing and Analysis Consortium (DPAC, https://www.cosmos.esa.int/web/gaia/dpac/consortium). Funding for the DPAC has been provided by national institutions, in particular, the institutions participating in the Gaia Multilateral Agreement.

\end{acknowledgments}

\begin{contribution}
Responsibilities for this work break down as:
PBK and MNI devised and developed the project, while PBK designed and performed the analysis and wrote the manuscript. TK and IO were involved in discussing and interpreting the results and editing the manuscript.


\end{contribution}

%
\facilities{Gaia, CFHT: MegaCam.}

\software{\textsc{astropy} \citep{2013A&A...558A..33A,2018AJ....156..123A}, \textsc{astroquery} \citep{2019AJ....157...98G}, \textsc{Gala} \citep{gala,adrian_price_whelan_2020_4159870}, \textsc{galpy} \citep[][]{2015ApJS..216...29B}, \textsc{matplotlib} \citep{Hunter:2007}, \textsc{numpy} \citep{2011CSE....13b..22V}, \textsc{scipy} \citep{2020SciPy-NMeth}, \textsc{sklearn} \citep[][]{scikit-learn}, \textsc{Ultranest} \citep{2014A&A...564A.125B,2019PASP..131j8005B,2021JOSS....6.3001B}.
 }


\appendix

\section{Incomplete fields of view}\label{sec:GC_removed}
Before exploring the distribution of the high-probability stars in the GC peripheries, some limitations on the data need to be addressed.  First, despite the quality cuts on the data undertaken in section \ref{sec:The_Data}, PGS does inherit some data artefacts from Gaia. Specifically, Gaia's scanning pattern is reflected in PGS through quality cuts and overall data availability. The effects of the less scanning in regions of the sky correlate to poor astrometric and photometric solutions, directly affecting the accuracy of the reported solutions. As a result, selecting the most reliable data can cause patterns of incompleteness across the field of view, and this affects our ability to explore the periphery for consistent extended tidal structure. Though we do not rely on the spatial distribution for the identification of GC members in our technique, we can still explore the distribution of those stars in ($x, y$) space and compare them to any known stream or tidal structure. Four GCs whose GC had plenty of stars but suffer from incomplete scanning, and those are NGC 5897, NGC 5904, NGC 6681 and NGC 6809. Secondly, the creation of the 2D surface density distributions described in section \ref{sec:Methods} can be subjected to over-smoothing if a GC has little or no high probability stars outside the tidal radius, either completely smoothing out any coherent structure, or there is no extended structure at all. Therefore, we adopted a criterion as to whether a GC is suitable to have its 2D smoothed density map analysed. For stars located outside the tidal radius, the sum of their probabilities must be larger than five, $\sum{\mathrm{P_{mem}}}>5$. Four GCs did not make this cut: NGC 288, NGC 2298, NGC 4590 and NGC 5466. Despite these shortcomings, we decided to include these eight clusters in our sample, but we can not confirm or exclude the existence of extended structure around them in PGS.


\begin{figure*}
\figurenum{10}
    \centering
    \includegraphics[width=0.9\columnwidth]{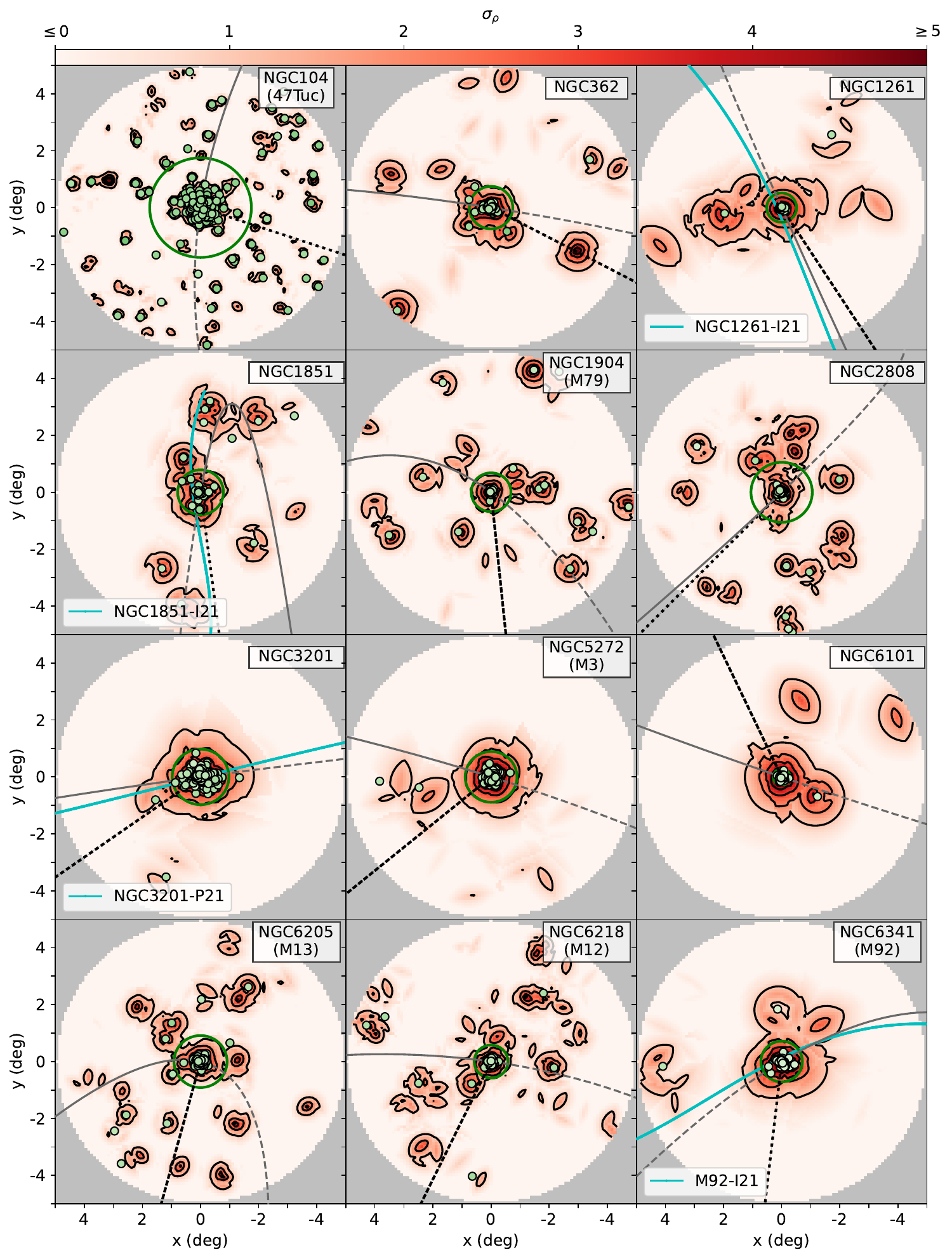}
    \caption{The same as Fig. \ref{fig:1_part1} but with the addition of the conservative selection of stars denoted by the green points. \textit{Stream track labels: I21 - \citet[][]{2021ApJ...914..123I}, P21 - \citet[][]{2021MNRAS.504.2727P}.}}
    \label{fig:1_part1_cons}
\end{figure*}

\begin{figure*}
\figurenum{10}
    \centering
    \ContinuedFloat
    \includegraphics[width=0.9\columnwidth]{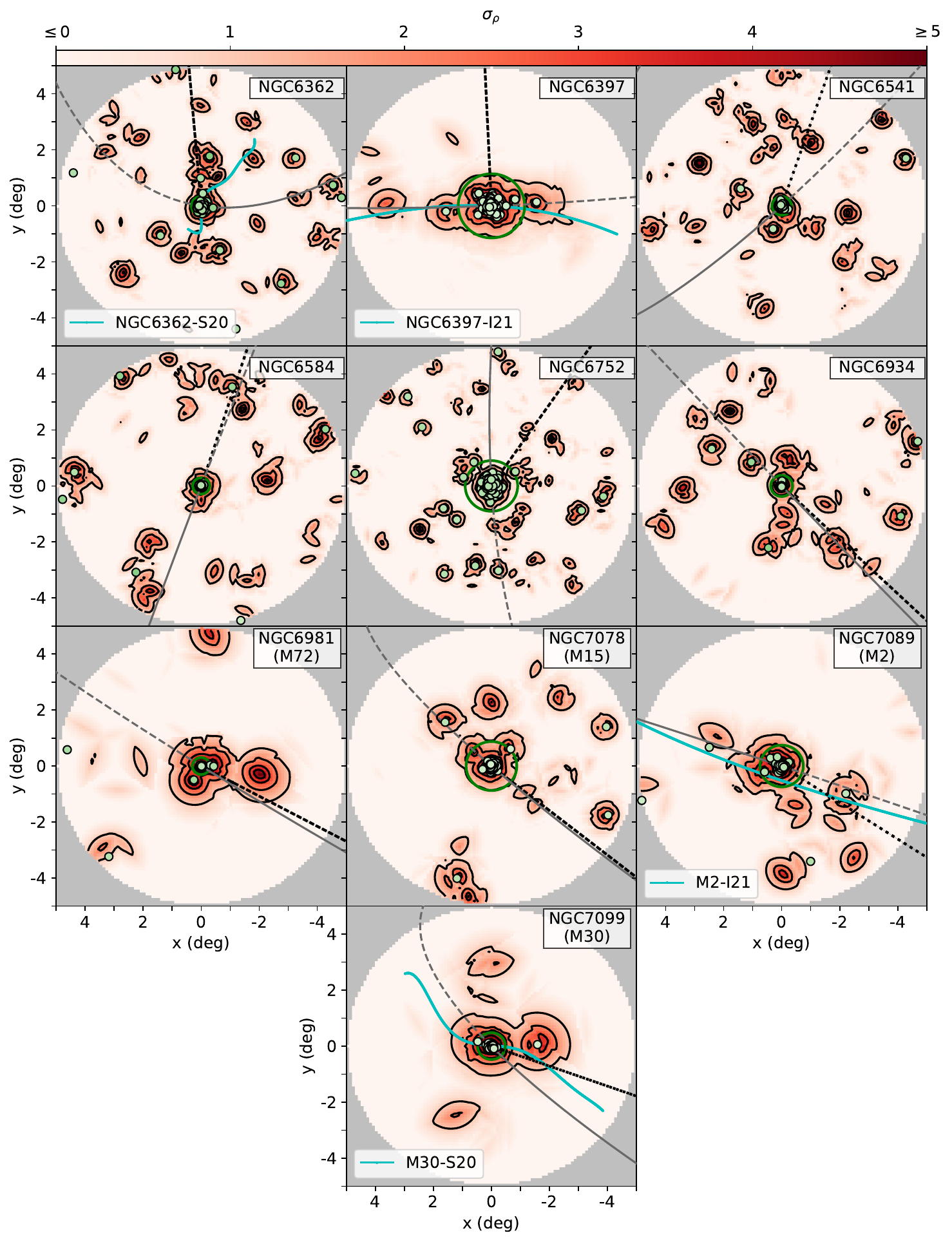}
    \caption[]{A continuation of Fig.  \ref{fig:1_part1_cons}. \textit{Stream track labels: S20 - \citet[][]{2020MNRAS.495.2222S}, I21 - \citet[][]{2021ApJ...914..123I}.} }
    \label{fig:1_part2_cons}
\end{figure*}

 \section{Additional Plots}

 \begin{figure*}
 \figurenum{11}
     \centering
     \includegraphics[width=0.9\columnwidth]{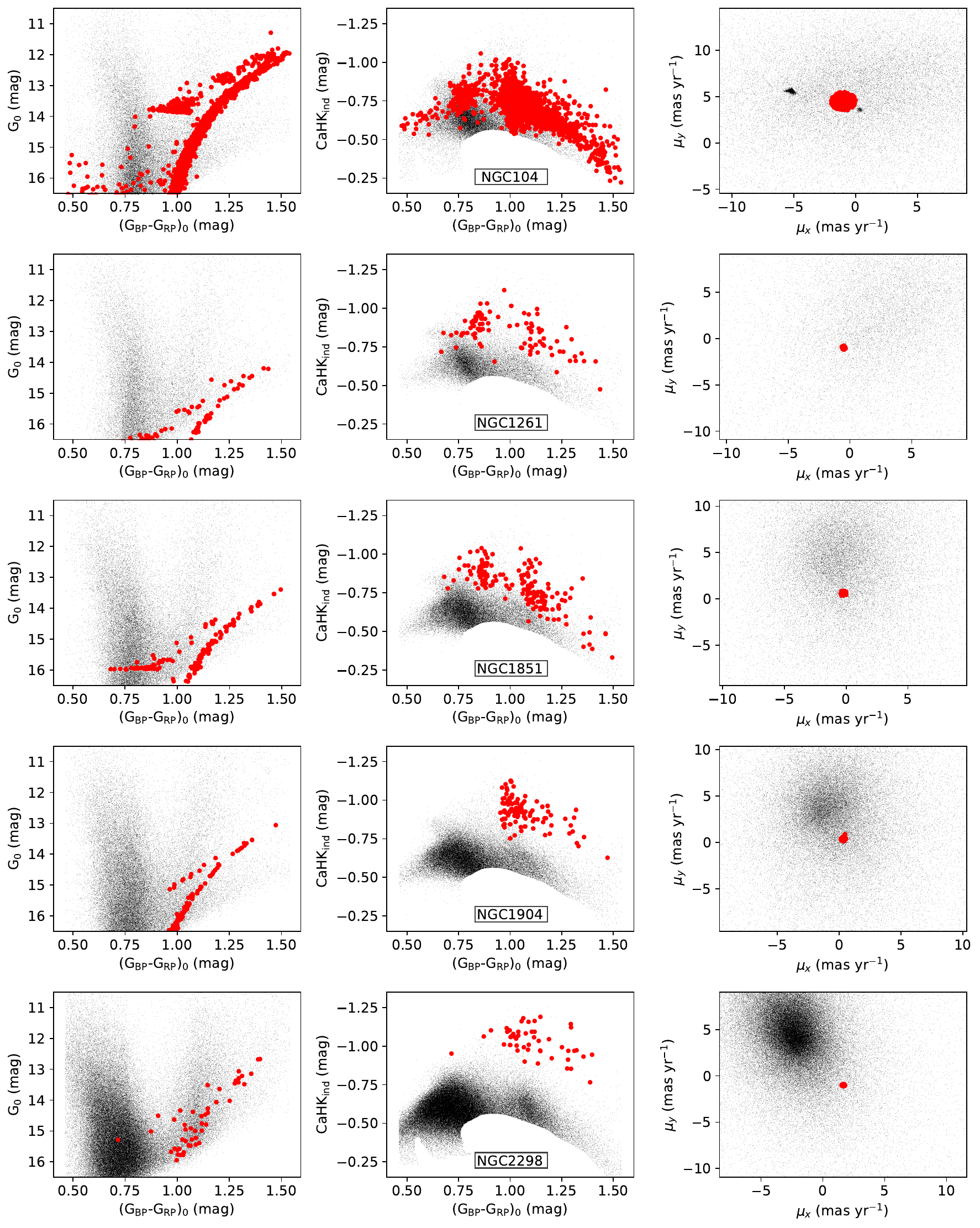}
     \caption{Extended version of Fig. \ref{fig:CMD}.}
     \label{fig:AP_1_part1_cons}
 \end{figure*}

 \begin{figure*}
     \centering
      \figurenum{11}
     \ContinuedFloat
     \includegraphics[width=0.9\columnwidth]{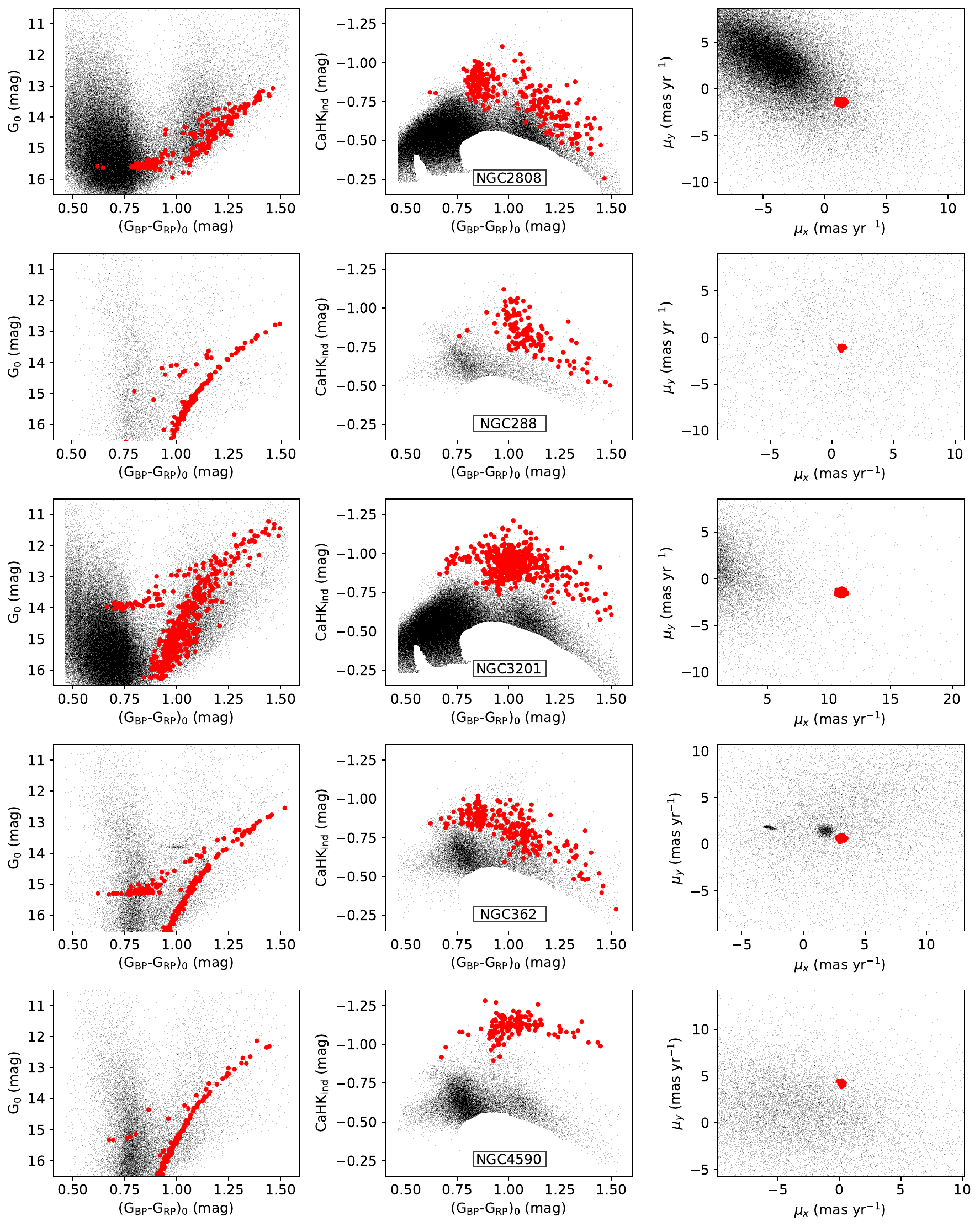}
     \caption[]{Extended version of Fig. \ref{fig:CMD}.}
     \label{fig:AP_1_part2_cons}
 \end{figure*}

 \begin{figure*}
     \centering
      \figurenum{11}
     \ContinuedFloat
     \includegraphics[width=0.9\columnwidth]{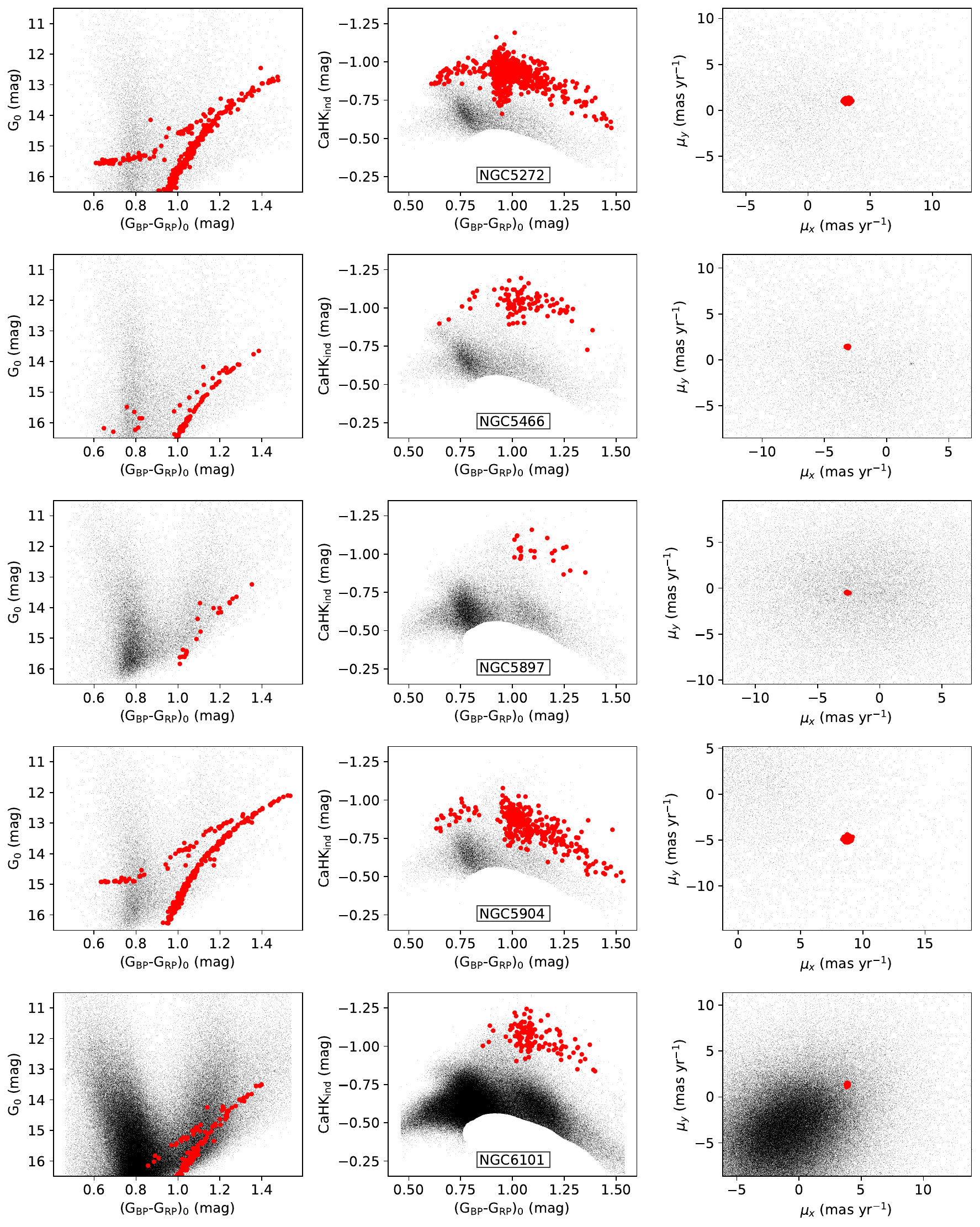}
     \caption[]{Extended version of Fig. \ref{fig:CMD}.}
     \label{fig:AP_1_part3_cons}
 \end{figure*}

 \begin{figure*}
     \centering
      \figurenum{11}
     \ContinuedFloat
     \includegraphics[width=0.9\columnwidth]{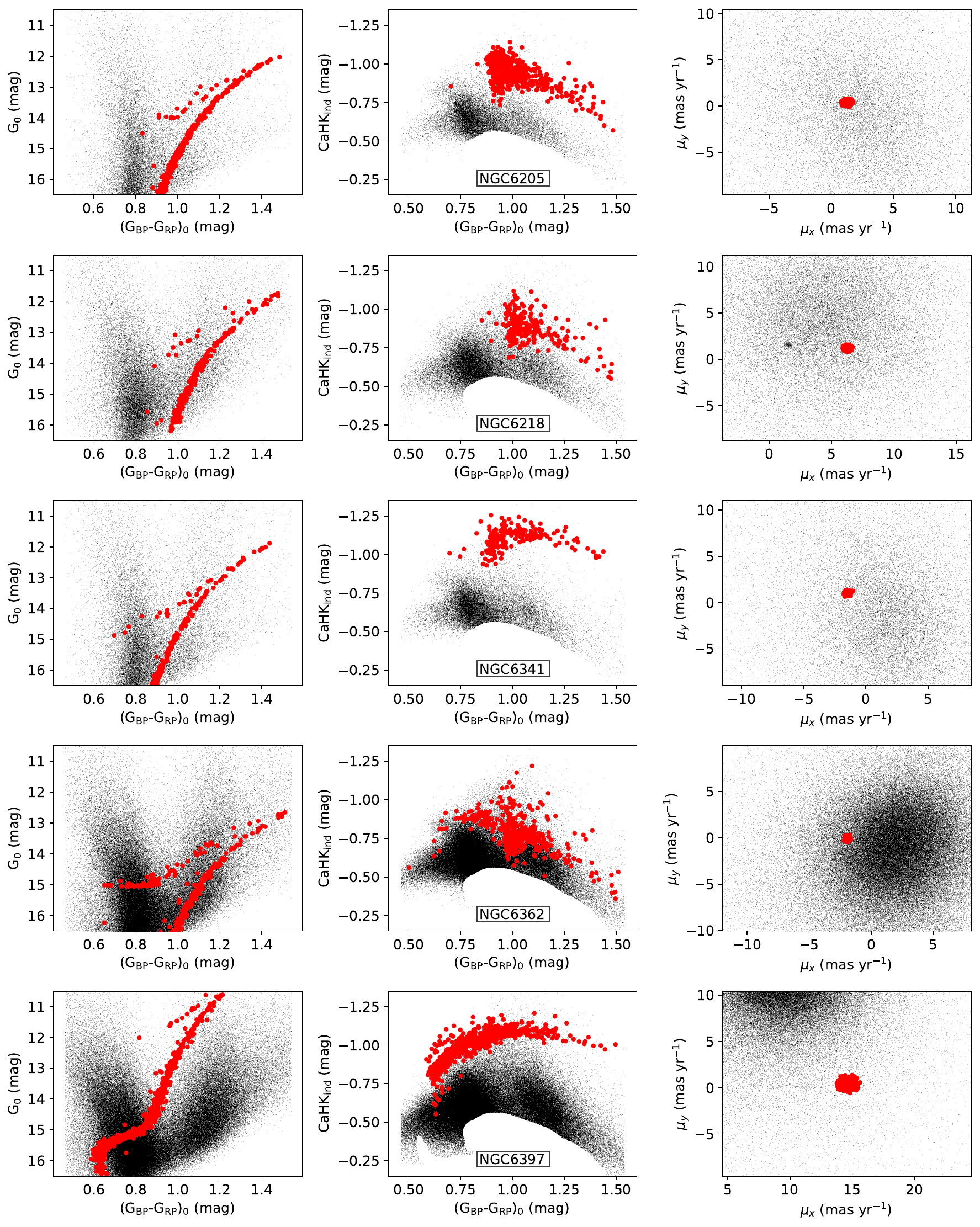}
     \caption[]{Extended version of Fig. \ref{fig:CMD}.}
     \label{fig:AP_1_part4_cons}
 \end{figure*}

 \begin{figure*}
     \centering
      \figurenum{11}
     \ContinuedFloat
     \includegraphics[width=0.9\columnwidth]{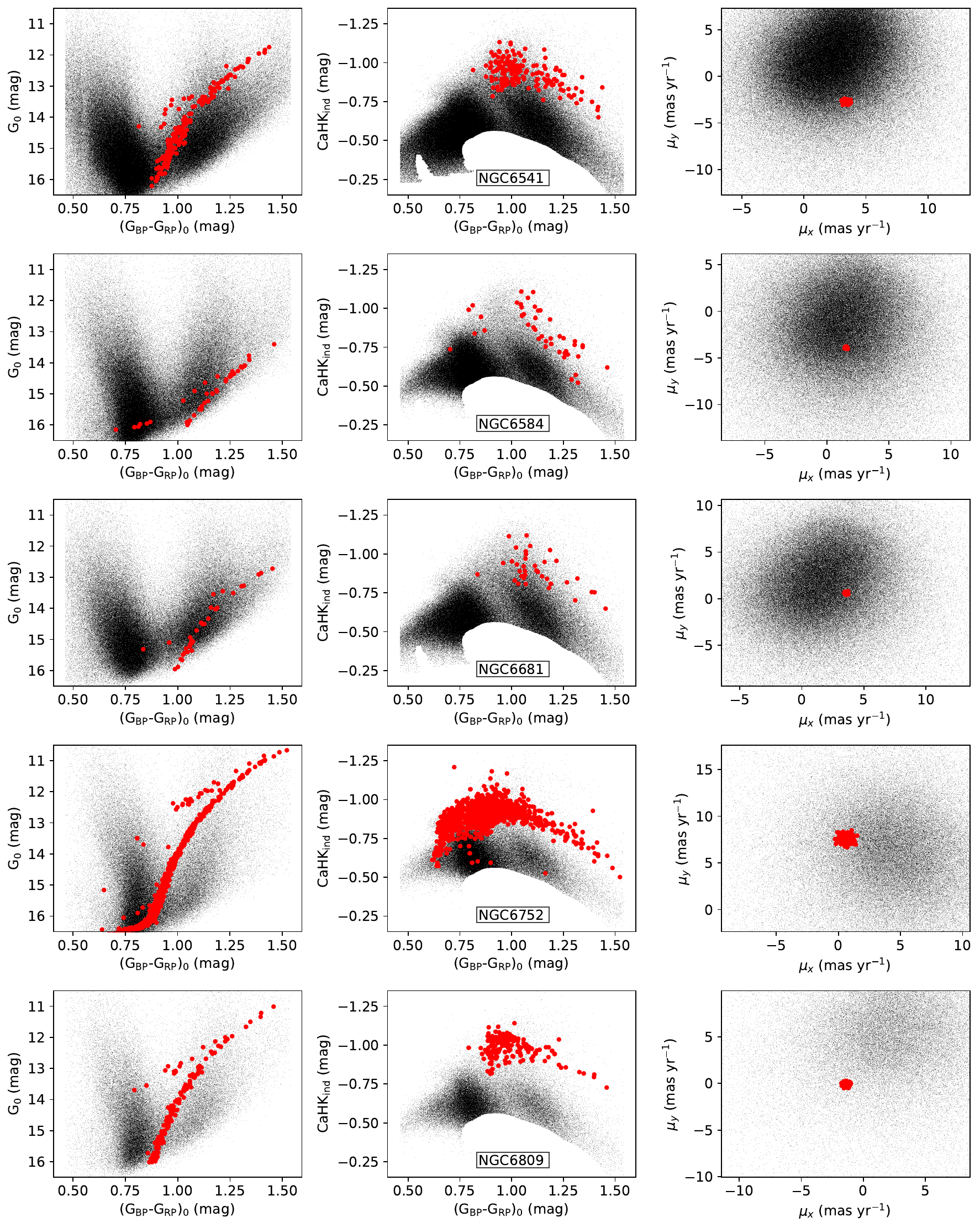}
     \caption[]{Extended version of Fig. \ref{fig:CMD}.}
     \label{fig:AP_1_part5_cons}
 \end{figure*}

 \begin{figure*}
     \centering
      \figurenum{11}
     \ContinuedFloat
     \includegraphics[width=0.9\columnwidth]{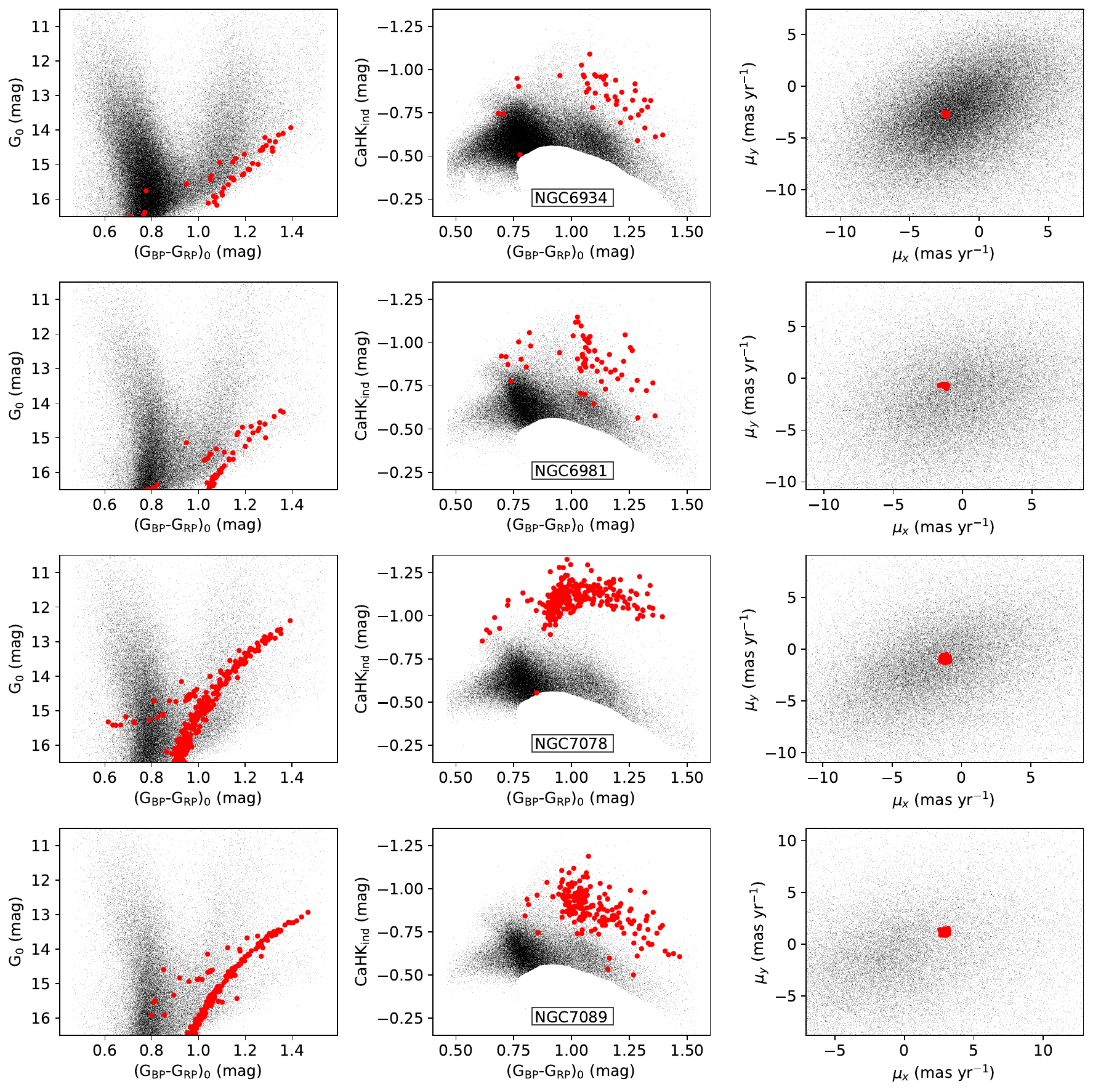}
     \caption[]{Extended version of Fig. \ref{fig:CMD}.}
     \label{fig:AP_1_part5_cons}
 \end{figure*}

\bibliography{Library_bibtex}
\bibliographystyle{aasjournalv7}



\end{document}